\documentclass[]{article}
\usepackage[T1]{fontenc}       
\usepackage{graphicx}
\usepackage{bm}
\usepackage{float}
\usepackage{hyperref}
\usepackage{verbatim}
\usepackage{graphicx}
\usepackage{dcolumn}
\usepackage{bm}
\usepackage{float}
\usepackage{hyperref}
\usepackage{fancyhdr}
\usepackage[nottoc]{tocbibind}
\usepackage{chngcntr}
\usepackage{authblk}
\usepackage[letterpaper, portrait, margin=1in]{geometry}
\usepackage[sort&compress,numbers,super]{natbib}
\usepackage{caption}
\usepackage{subcaption}
\usepackage[utf8]{inputenc}
\usepackage{color}
\usepackage{mathtools}
\usepackage{comment}
\usepackage[T1]{fontenc}
\usepackage{lmodern}

\usepackage{blindtext}

\usepackage{subfiles}

\title{\textbf{Non steady-state thermometry with optical diffraction tomography}}
\author[1]{Adarsh B Vasista}
\author[1,2]{Bernard Ciraulo}
\author[1,*]{Jaime Ortega Arroyo}
\author[1,*]{Romain Quidant}
\affil[1]{\textit{Nanophotonic Systems Laboratory,Department of Mechanical and Process Engineering, ETH Zürich, 8092 Zürich, Switzerland }}
\affil[2]{\textit{Pediatric Molecular Neuro-Oncology Research, University Children's Hospital Zürich, Balgrist Campus, 8008 Zürich, Switzerland}}

\affil[*]{\textit{\textbf{E-mail:} jarroyo@ethz.ch; rquidant@ethz.ch}}

\date{}

\begin{document}


\maketitle



\begin{abstract}
Measurement of local temperature using \textit{label-free} optical methods has gained importance as a pivotal tool in both fundamental and applied research. Yet, most of these approaches are limited to steady-state measurements of planar heat sources. However, the time taken to reach steady-state is a complex function of the volume of the heated system, the size of the heat source, and the thermal conductivity of the surroundings. As such, said time can be significantly longer than expected and many relevant systems involve 3D heat sources, thus compromising reliable temperature retrieval. Here, we systematically study the thermal landscape in a model system consisting of optically excited gold nanorods (AuNRs) in a microchamber using  optical diffraction tomography (ODT) thermometry. We experimentally unravel the effect of thermal conductivity of the surroundings, microchamber height, and pump pulse duration on the thermodynamics of the microchamber. We benchmark our experimental observations against 2D numerical sumulations and quantitative phase imaging (QPI) thermometry. We also demonstrate the advantage of ODT thermometry by measuring thermal landscapes inaccessible by QPI thermometry in the form of non-planar heat sources embedded in complex environments such as biological cells. Finally, we apply ODT thermometry to a complex dynamic system consisting of colloidal AuNRs in a microchamber.    

\end{abstract}

\maketitle
\section{Introduction}
Measuring temperature reliably at the nano- and micro-scale is not only key to answering fundamental thermodynamic questions at these scales, but also in a variety of applications like photothermal cancer therapy \cite{1,2,3}, drug delivery \cite{4}, photocatalysis\cite{5,6}, thermal lensing \cite{7,8}, microfluidics \cite{16,42,43,46}, vibrational spectroscopy using mid-infrared photothermal microscopy\cite{34,35,36} etc. Nonetheless, the non-propagative nature of heat poses a challenge to reliably and accurately measure temperature at these scales, especially in non steady-state conditions.

Various optical thermometry techniques have recently emerged to address this need, and we can broadly categorize them into \textit{label-based} and \textit{label-free methods}. The working principle for label-based methods relies on measuring a temperature sensitive emission signature such as Raman scattering\cite{9,10}, fluorescence anisotropy\cite{11}, fluorescence intensity\cite{12,13}, fluorescence spectra\cite{14,15}, photoluminescence life time\cite{45} from a set of molecular probes. While these methods can measure temperature in non steady-state conditions, they face drawbacks like slow read-out rates\cite{14}, low sensitivity\cite{17}, lack of reliability\cite{12,13}, and most importantly the need to place the molecular probes in the system, which is not always feasible. To circumvent these issues, label free methods such as infra-red imaging\cite{18}, X ray absorption spectroscopy\cite{19},quantitative phase imaging (QPI)\cite{20}, and mid infrared photothermal microscopy \cite{34,35,36} have been proposed. Out of all label-free approaches QPI is one of the most promising ones, due to its ease of implementation into commercial microscopes, its speed and high resolution temperature retrieval. 

QPI thermometry is based on measuring the optical path differences of a probe beam as a result of the small  temperature-induced changes in the refractive index of a material. Several different implementations of QPI exist, either in the form of inline\cite{40}, off-axis\cite{16,46} or shearing-based holography\cite{20,41}, yet all extract an optical path length difference from the measured phase change. Nonetheless, retrieving temperature profiles from these optical path length changes relies on algorithms that assume the system is in steady-state and the temperature profile follows a $\frac{1}{r}$ decay from a planar heat source, where $r$ is the radial coordinate\cite{20}. Unfortunately, these assumptions restrict the range of systems to which QPI-thermometry can be applied.   

As a promising alternative to QPI-thermometry, ODT has been widely used to accurately determine 3D refractive index maps of biological systems \cite{22,23,24,26,27,29,30}, to study material anisotropy \cite{31}, and to perform vibrational spectroscopy based on mid-infrared photothermal microscopy\cite{35}. Recently, ODT-based thermometry has been experimentally demonstrated in the steady-state\cite{28} without the need of any assumptions other than a look-up-table that relates the measured 3D refractive index change from a thermo-optical material to a temperature change. As a result, this technique has the potential to study more complex temperature-dependent phenomena inaccessible to QPI, namely systems that do not satisfy the steady-state and  non-planar heat source assumptions. 

In this work, we systematically studied, experimentally and numerically, a non steady-state system in the form of a microchamber undergoing photothermal conversion by gold nanorods to assess the performance of ODT based thermometry. We specifically use QPI to benchmark the conditions that push the system away from steady-state and to identify the mechanism responsible for such. We find that for our system heat accumulation extends the time to reach steady-state, and can be tuned by either the height of the microchamber or the thermal conductivity of the surroundings. Under non steady-state conditions we validate ODT thermometry against simulations, and showcase that QPI despite accurately retrieving the temperature gradient, it underestimates the absolute value of the local temperature. Finally, we apply ODT thermometry to three representative non-steady state systems, where photothermal conversion is achieved by non-planar heat sources in the form of: colloidal gold nanoparticles freely diffusing in aqueous media, of nanoparticle clusters embedded in a 3D hydrogel, and of AuNRs internalised by cells. As such our work highlights a promising approach to address the knowledge gap of heat propagation and thermometry at the nano and micro length scales. 
%


\section{Results and discussions}
\subsection{Working principle}
    \begin{figure*}[h!]
    \centering
    \includegraphics[width=\linewidth]{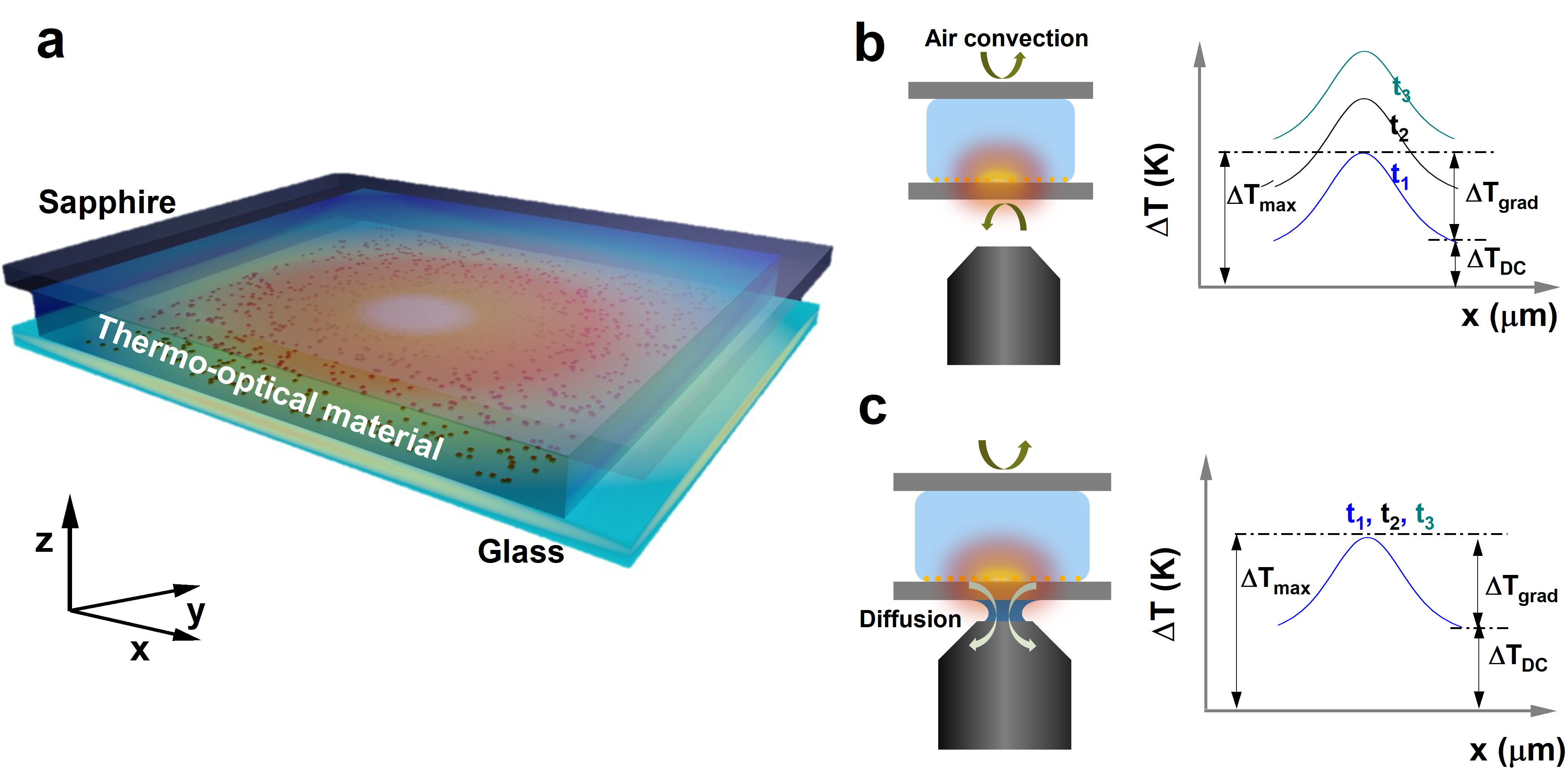}
    \caption{\textit{The sample and the technique.} (a) Schematic representation of the microchamber sample formed by a gold nanorod (AuNR) functionalised glass substrate and sapphire superstrate separated by a silicone spacer. The chamber is filled with water as the thermo-optical material with a known refractive index.  A 785 nm laser beam resonantly excites the AuNRs, causing the medium surrounding the AuNRs to heat up and change its refractive index. A probe laser at 465 nm measures the phase shift due to the altered refractive index. (b,c) Schematics representing the phenomena of heat transfer in the microchamber when the chamber is probed using air-immersion and oil-immersion objectives respectively and its effect on the temporal evolution of the thermal profiles. In the case of the air objective, the primary mechanism of heat transfer is natural air convection resulting in the continuous increase of the thermal floor ($\Delta T_{DC}$) unlike the oil-immersion case where the immersion oil acts as a thermal bridge between the chamber and the metallic case of the objective lens.  }
    \label{fig:Fig1}
\end{figure*}

 Figure \ref{fig:Fig1} depicts the working principle of the experiment. As nano sources of heat, we used AuNRs immobilised on a glass substrate (S1 methods). A microfluidic chamber was, then, prepared by sandwiching a thermo-optical material, water in our case, between the glass substrate and a sapphire superstrate  (Figure \ref{fig:Fig1}a). A pump beam of wavelength close to the absorption maximum of the AuNRs (785 nm) excites the substrate and heats the sample. This in turn changes the temperature-dependent refractive index of water, thereby spatially encoding the thermal profile in the form of wavefront changes. Measuring these temperature-induced wavefront changes  form the basis of either phase-based temperature techniques: QPI and ODT thermometry. 

 When the AuNRs heated using a pump laser, we can define two characteristic time scales defining the thermodynamic state of the system. (i) The timescale to reach the local steady-state in the immediate vicinity of the heated nanorods. This timescale can be understood as the time for the thermal gradient in the sample to establish and is of the order of a few ms for a typical beam size of $\sim10 \mu$m. (ii) The timescale for the entire microchamber system to reach steady-state. This parameter is a complex function of the volume of the thermo-optical material, translating to the chamber height, and the thermal conductivity of the surroundings. 

If the microchamber is not in thermal contact with a heat sink, then the chamber thermalizes with the outside environment through natural air convection, as shown in figure \ref{fig:Fig1} (b). In this case, local thermalization happens quickly, setting up the thermal gradient. However, the lower efficiency of the natural air convection limits the dynamics of the heat transfer and results in the build-up of heat inside the microchamber. This heat build-up in the chamber manifests as a constant increase in the temperature profile. Hence, if one monitors the temporal evolution of the thermal profile, though the microchamber reaches a local steady-state quickly, establishing the thermal gradient ($\Delta T_{grad}$), it experiences a continuous increase of the thermal floor ($\Delta T_{DC}$). 

On the other hand, the thermodynamics of the heated microchamber can be drastically altered by providing access to a thermal sink by substituting the air-immersion objective with an oil-immersion one. In such a case, the refractive index matching oil acts as a thermal bridge between the microchamber and the metallic body of the objective lens and its respective optomechanical elements, as shown in figure \ref{fig:Fig1} (c). As the microchamber is in thermal contact with a large thermal sink, the time taken to reach the global steady-state is considerably less, thus avoiding the continuous increase of the thermal floor ($\Delta T_{DC}$).

\begin{figure*}[h!]
    \centering
    \includegraphics[width=\linewidth]{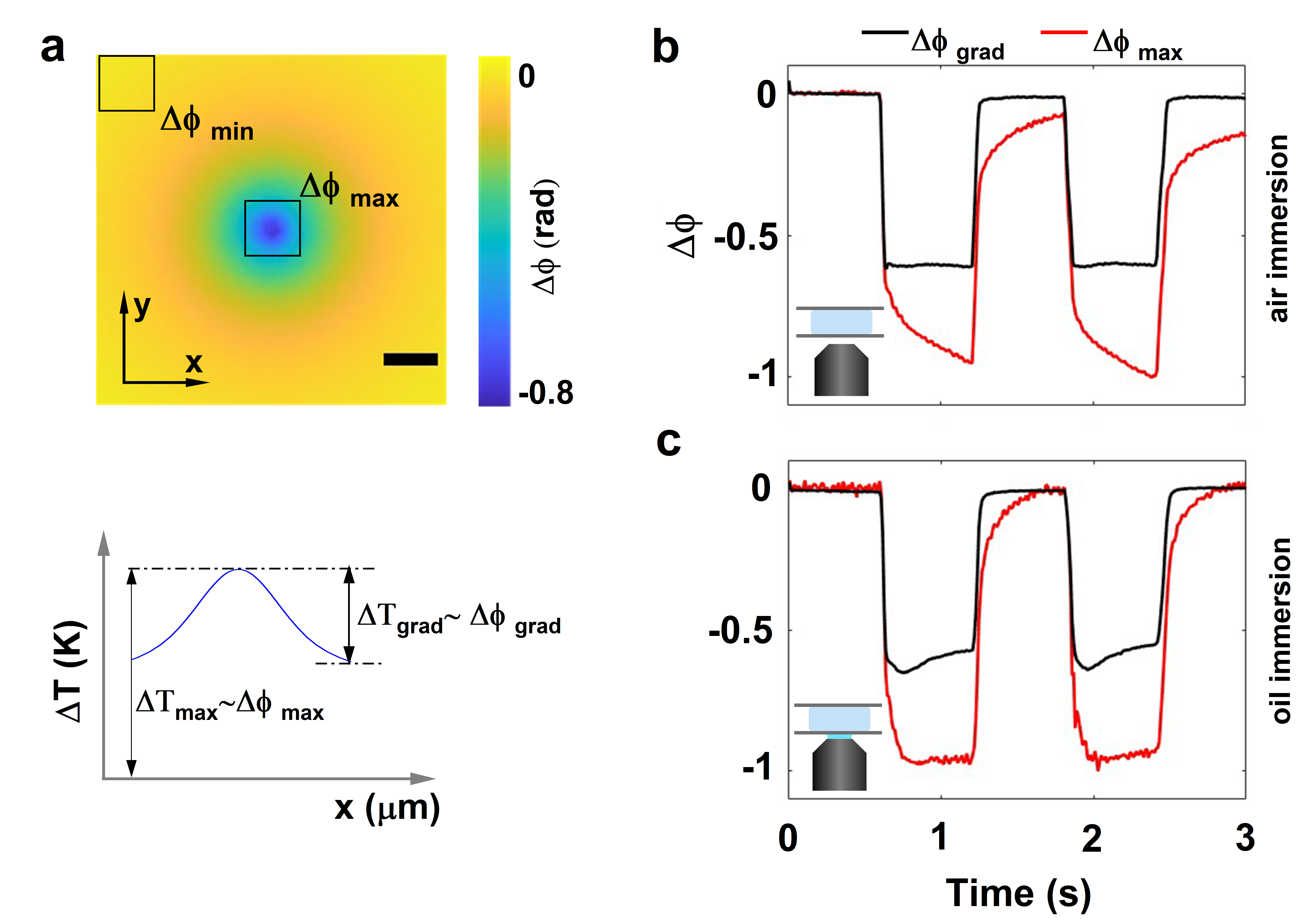}
    \caption{\textit{Assessing steady-state dynamics by phase imaging.}  (a) (\textit{top}) A representative phase difference image measured by subtracting the phase image with the heating laser switched ON from a reference phase image measured with heating laser switched OFF and (\textit{bottom}) its corresponding thermal profile properties. The maximum and minimum phase shift within the field of view, acquired due to the heating of nanorods, is termed $\Delta\phi_{max}$ and $\Delta\phi_{min}$, respectively. The difference between the $\Delta\phi_{max}$ and $\Delta\phi_{min}$ determines the phase gradient ($\Delta\phi_{grad}$) of the phase image. The scale bar is 15 $\mu$m. The maximum phase shift accumulated ($\Delta\phi_{max}$) corresponds, in a closed system, to the absolute increase of the temperature ($\Delta T_{max}$) and the phase gradient ($\Delta\phi_{grad}$) corresponds to the thermal gradient in the image ($\Delta T_{grad}$).  (b,c) Time evolution of $\Delta\phi_{grad}$ and $\Delta\phi_{max}$ for a chamber height of 500 $\mu$m probed using air and oil immersion objectives respectively. } 
    \label{fig:Fig2}
\end{figure*}

\subsection{Time evolution of phase maps}
To understand the thermodynamics of the microchamber, we study the temporal evolution of phase difference maps with pump-probe phase imaging in an off-axis holography configuration (Figure \ref{fig:Fig2}, SI methods). The AuNRs in the microchamber were heated upon irradiation with a time modulated pump laser of wavelength 785 nm (close to the absorption maximum of the nanorods). The optical path difference, OPD, due to optical pumping and thereby local heating was measured by the difference between pump ON and pump OFF states of a probe beam with wavelength 465 nm \cite{16}.   

 For quasi-infinite systems, and at $r=0$ the steady-state occurs on a timescale given by $\tau \sim \frac{D^2}{4a_{s}}$ \cite{7}, where $D$ is the diameter of the heat source and $a_{s}$ is the thermal diffusivity of the medium. In the case of photothermal conversion of nanoparticle ensembles, i.e. this experiment, the diameter of the heat source corresponds to the size of the pump beam. Therefore, for a pump beam size of 10 $\mu$m, $\tau  \sim$ 175 $\mu$s. However, this timescale does not hold for positions away from the centre of the heat source nor for finite systems such as a microchamber. Instead the time to reach steady-state is a complex function of: the height of the chamber, the thermal conductivities of the substrate and the superstrate, the position away from the heat source, and the heat transfer properties of the thermal sink. 
 

To follow the thermal dynamics of the system, we tracked the temporal evolution of the phase of the probe beam, which is a suitable metric for finite real-world systems given the relation between OPD and temperature. Figure \ref{fig:Fig2}(a) shows a representative phase difference map between pump ON (hot) and pump OFF(cold) states, where a significant phase dip at the center (laser excitation spot) is observed, as expected due to the negative thermo-optical coefficient of water. To understand the temporal phase response, we defined two important parameters: (i) $\Delta\phi_{max}$ - the maximum phase shift acquired by the probe beam due to heating and (ii) $\Delta\phi_{grad}$ - the maximum phase gradient in the image calculated by subtracting the phase difference value at the edge of the phase image ($\Delta\phi_{min}$) from the phase change induced at the center of the image due to heating. Intuitively, $\Delta\phi_{max}$ and $\Delta\phi_{grad}$ report on the absolute temperature change and thermal gradient in the microchamber, respectively as represented by figure \ref{fig:Fig2} (a). For instance at steady-state, when the temperature no longer changes within the sample, we expect the phase to converge to a constant value.

To understand the temperature dynamics of the mircochamber as a function of the thermal conductivity of the surroundings, we followed the evolution of $\Delta\phi_{max}$ and $\Delta\phi_{grad}$ in a microchamber of height 500 $\mu$m probed in two different configurations: in the absence (air objective) and presence (oil-immersion objective) of a thermal sink (figures \ref{fig:Fig2}(b) and (c)). In the absence of a thermal sink, measurements with an air objective (figure \ref{fig:Fig2}(b)), the dynamics of the $\Delta\phi_{max}$ did not saturate within the timescale of the experiment, whereas, $\Delta\phi_{grad}$ saturated shortly after the pump pulse was switched ON. This indicates on the one hand that the system had not reached a steady-state. On the other hand, even though the system was not in steady-state, the shape of the thermal profile remained the same, as indicated by the stabilization of the phase gradient. 

Figure \ref{fig:Fig2}(c) shows the time evolution of $\Delta\phi_{max}$ and  $\Delta\phi_{grad}$ of the microchamber in the presence of a thermal sink by using an oil immersion objective lens. Here $\Delta\phi_{max}$ saturated within the timescale of the experiment, while the $\Delta\phi_{grad}$ showed oscillatory behaviour before saturating. We suggest that the oscillatory behaviour in the $\Delta\phi_{grad}$ may result from the interplay between the different timescales involved as heat diffuses across  multiple finite-sized materials with different thermal conductivities. For instance, the large thermal conductivity of the metal from the objective accentuates the diffusion along the axial direction, resulting in the short-lived oscillatory behaviour in the $\Delta\phi_{grad}$ (figure S8, SI).  Comparing figures \ref{fig:Fig2} (b) and (c) we can conclude that coupling the system to a heat sink via the oil immersion modifies the thermal dynamics by specifically pushing the system towards the steady-steady much faster compared to the air immersion configuration. 

\begin{figure*}[h!]
    \centering
    \includegraphics[width=\linewidth]{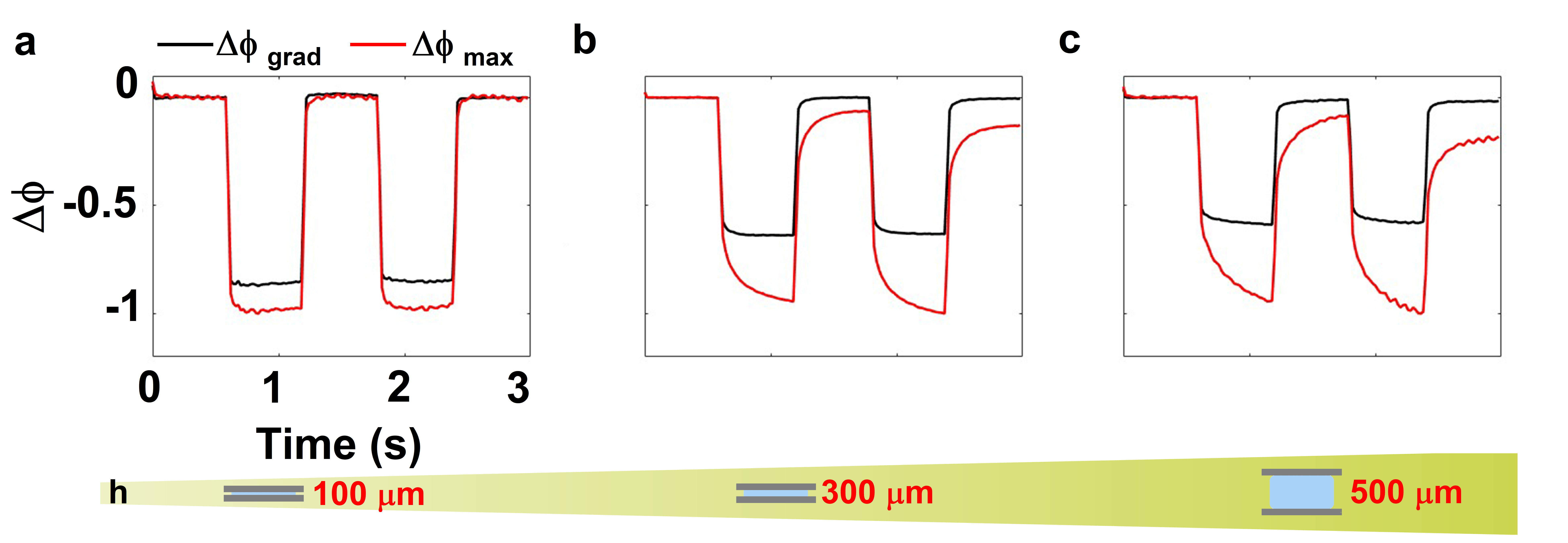}
    \caption{\textit{Assessing steady-state dynamics by phase imaging.} (a)-(c) show the time evolution of the phase gradient $\Delta\phi_{grad}$ and maximum phase shift accumulated $\Delta\phi_{max}$ for chamber heights of 100 $\mu$m, 300 $\mu$m, and 500 $\mu$m probed using air objective respectively.  } 
    \label{fig:Fig3}
\end{figure*}

Further, to understand the effect of height of the microchamber on the temperature dynamics, we studied the temporal evolution of phase for three different chamber heights at a fixed pump beam size and power in the air immersion configuration (figures \ref{fig:Fig3} (a)-(c)). For the 100 $\mu$m chamber, the dynamics of $\Delta\phi_{max}$ and $\Delta\phi_{grad}$ saturated within the time scales of the experiment, suggesting that the system had reached steady-state within the pulse duration of pump beam (600 ms). Upon increasing the chamber height to either 300 $\mu$m or 500 $\mu$m, $\Delta\phi_{max}$ no longer saturated,  whereas  $\Delta\phi_{grad}$ did so shortly after the pump pulse was switched ON. As the height of the microchamber was increased, so did the distance between the high thermal conductivity sapphire superstrate, which acts a heat sink, and the heat source thus affecting the thermodynamics of the chamber. These results, again, indicate that the system with increased chamber heights had not reached steady-state, however, the shape of the temperature profile remained the same, as indicated by the stabilisation of the thermal gradient. In other words, the difference between the temperature probed by $\Delta\phi_{max}$ and $\Delta\phi_{grad}$ corresponded to a uniform temperature shift , DC offset, within the imaged area. It is also important to note that the thermal relaxation dynamics (cooling) also critically depended on the chamber height, and the relaxation timescale was slower for larger chamber heights. Thus to probe non steady-state dynamics of the microchamber, we specifically used the air immersion objective lens configuration for the rest of the experiments detailed here. 

To study the effect of material properties of the superstrate (thermal conductivity, in particular) on temperature dynamics, we changed the superstrate of the microchambers from sapphire ($\kappa_{saph}$=30 W/mK) to glass ($\kappa_{glass}$=0.9 W/mK). Overall, the thermal conductivity of the superstrate had a minimal effect on the temperature dynamics for the chambers heights of 100 $\mu$m and 500 $\mu$m, and only showed a marginal effect when the chamber height was 300 $\mu$m (Figure S7 in SI). 

To further understand the complex relationship between the chamber height and the temporal dynamics of temperature, we performed 2D numerical simulations using COMSOL Multiphysics for a fixed pump spot size of 10 $\mu$m (SI Section S4). We calculated the temporal evolution of temperature in two extreme cases of the chamber height: 50 $\mu$m and 500 $\mu$m. Numerical simulations revealed that the time to reach steady-state for a 500 $\mu$m chamber was about 17 minutes, while for a 50 $\mu$m chamber it was about 7.5 minutes.  However, it should be noted that the conclusions drawn here are limited to the cases where natural air convection is the primary mechanism by which the microchamber interacts with the environment. Under such conditions, heat accumulates within the system and leads to a rise in the global temperature. 

Furthermore, we can conclude that the time to reach steady-state, in the case of microchambers and closed systems in general imaged using air objectives, significantly depends on the chamber height (translates to the volume of water) and is orders of magnitude higher than the expected theoretical value of 175 $\mu$s for a spot size of $\sim$10 $\mu$m. As such, this case highlights the overall need to exercise caution when estimating the steady-state dynamics of the system, as well as the need to take into account the system as a whole, including its surroundings to obtain an accurate picture of the underlying thermodynamics. However, it has to be noted that the conclusions drawn here applies to a closed microchambers with a finite height and can not be extended to quasi-infinite systems as $\Delta\phi_{max}$ diverges for an infinite system.

\begin{figure*}[h]
    \centering
    \includegraphics[width=\linewidth]{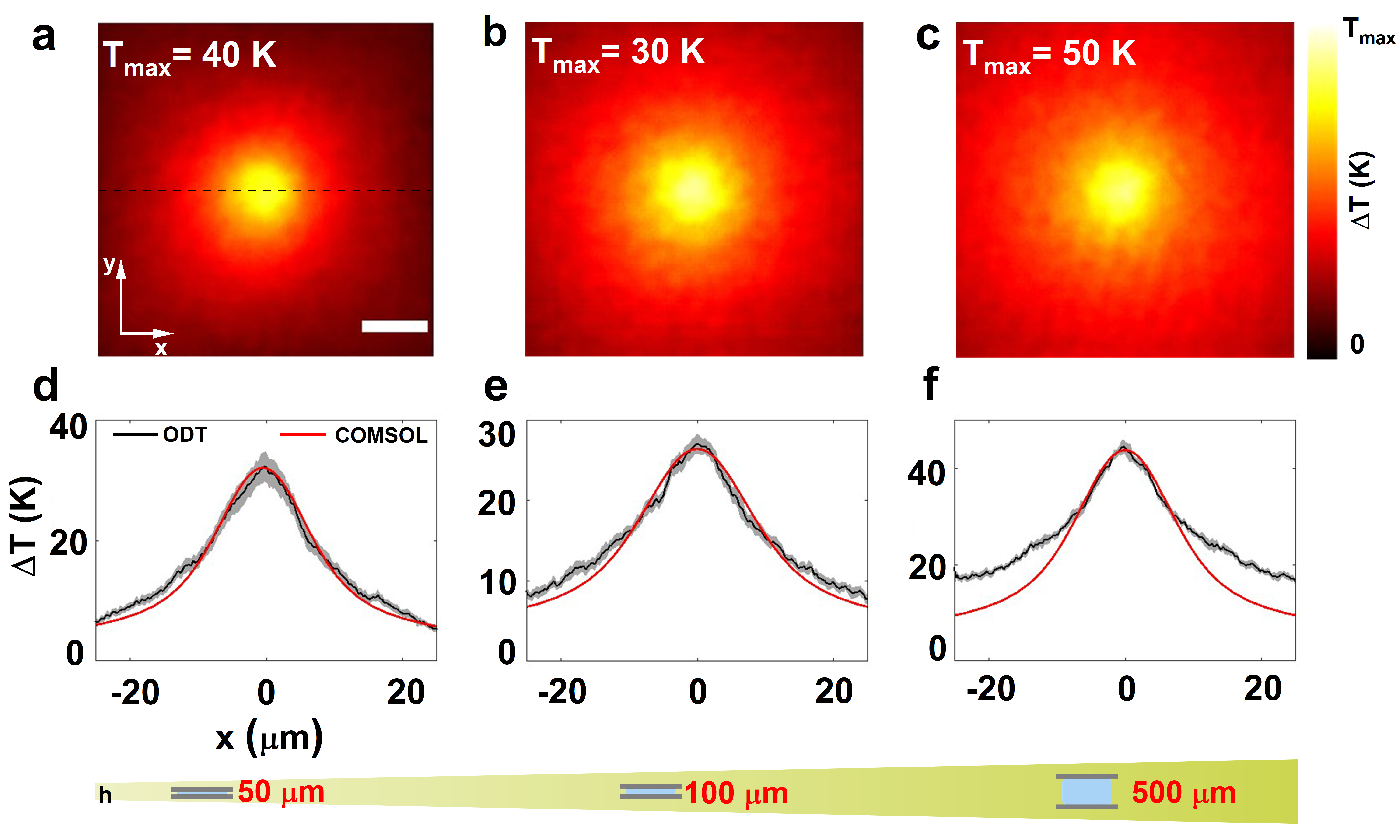}
    \caption{ \textit{Application of \textbf{ODT} thermometry to planar heat sources}. (a) - (c) Experimental thermal map at $Z=0$ measured using ODT for the chamber heights of 50 $\mu$m , 100$\mu$m, and 500 $\mu$m respectively. (d)-(f) Comparison of the line profile plotted across Y$=$0 (shown as a dashed line in (a)) with the numerically calculated thermal profile. The pump duration was fixed to 80 ms and the camera frame rate to 10 Hz. The scale bar is 5 $\mu$m.
} 
    \label{fig:Fig4}
\end{figure*}




\subsection{Non steady-state thermometry: planar heat source}

We first applied ODT thermometry to understand the thermal profiles of a planar heat source and systematically probed the microchambers in a pump-probe manner with three different chamber heights, keeping the beam size (12 $\mu$m) and pump pulse duration (80 ms) constant. In these microchambers, AuNRs were anchored to the glass substrate forming a planar heat source when excited by a pump laser, similar to those used to measure the phase maps in figure \ref{fig:Fig2}. Figures \ref{fig:Fig4} (a)-(c) show the cross section of measured temperature profile (at z=0) for chamber heights of 50 $\mu$m, 100 $\mu$m, and 500 $\mu$m respectively. Numerical simulations using COMSOL were carried out to corroborate the experimental data. To compare the experimental data with simulations, we plotted the line profile along y=0, represented as the dashed black line in figure \ref{fig:Fig4} (a). 

We found an excellent agreement between numerical simulations and experiments for chamber heights of 50 $\mu$m and 100 $\mu$m as shown by the line profiles in (figures \ref{fig:Fig4}(d) and (e) ). However, there is a mismatch for the 500 $\mu$m chamber, which we attributed to the slower relaxation dynamics. In detail, given that the measurements were performed in a pump-probe scheme there is an intrinsic assumption that the system cool sufficiently fast enough such that the pump OFF state does not have any residual heat left from the previous heating cycle. For the 80 ms pump duration, this condition was not satisfied, as the pump duration was comparable to the frame time of the camera (T$_{pmp}$=80 ms and T$_{cam}$=100 ms). Hence the residual heat in the system interfered with the measurement and appeared in the form of deviations from the theoretically expected $\frac{1}{r}$ profile. To verify this hypothesis we probed the 500 $\mu$m chamber with different pump pulse durations (5 ms, 20ms, and 80 ms) whilst keeping the beam size constant. As expected, the line profiles extracted for the shorter pump pulse duration of 5 ms and 20 ms matched well with the numerically calculated profiles (figure S9, SI). 

To establish the non steady-state nature of the temperature dynamics, we benchmarked the ODT measurements against QPI thermometry. As mentioned earlier, QPI thermometry, in its most general form, assumes steady-state and presupposes the $\frac{1}{r}$ decay in the temperature profile. The thermal profiles extracted using QPI match very well with that of ODT thermometry up to a DC shift due to the global heat accumulation as predicted by the temporal evolution of the phase gradient (figure \ref{fig:Fig2}, figure S10 SI). As expected, the value of the constant shift between the thermal profiles extracted from QPI and ODT depends on the chamber height.

Overall, by systematically studying the temperature profiles in microchambers with multiple chamber heights and pump durations and benchmarking the results with numerical simulations we show that ODT thermometry can be applied to study a wide class of non steady-state thermodynamic systems.

\subsection{ODT thermometry: Non-planar, spatially fixed heat sources}
\begin{figure*}[h!]
    \centering
    \includegraphics[width=\linewidth]{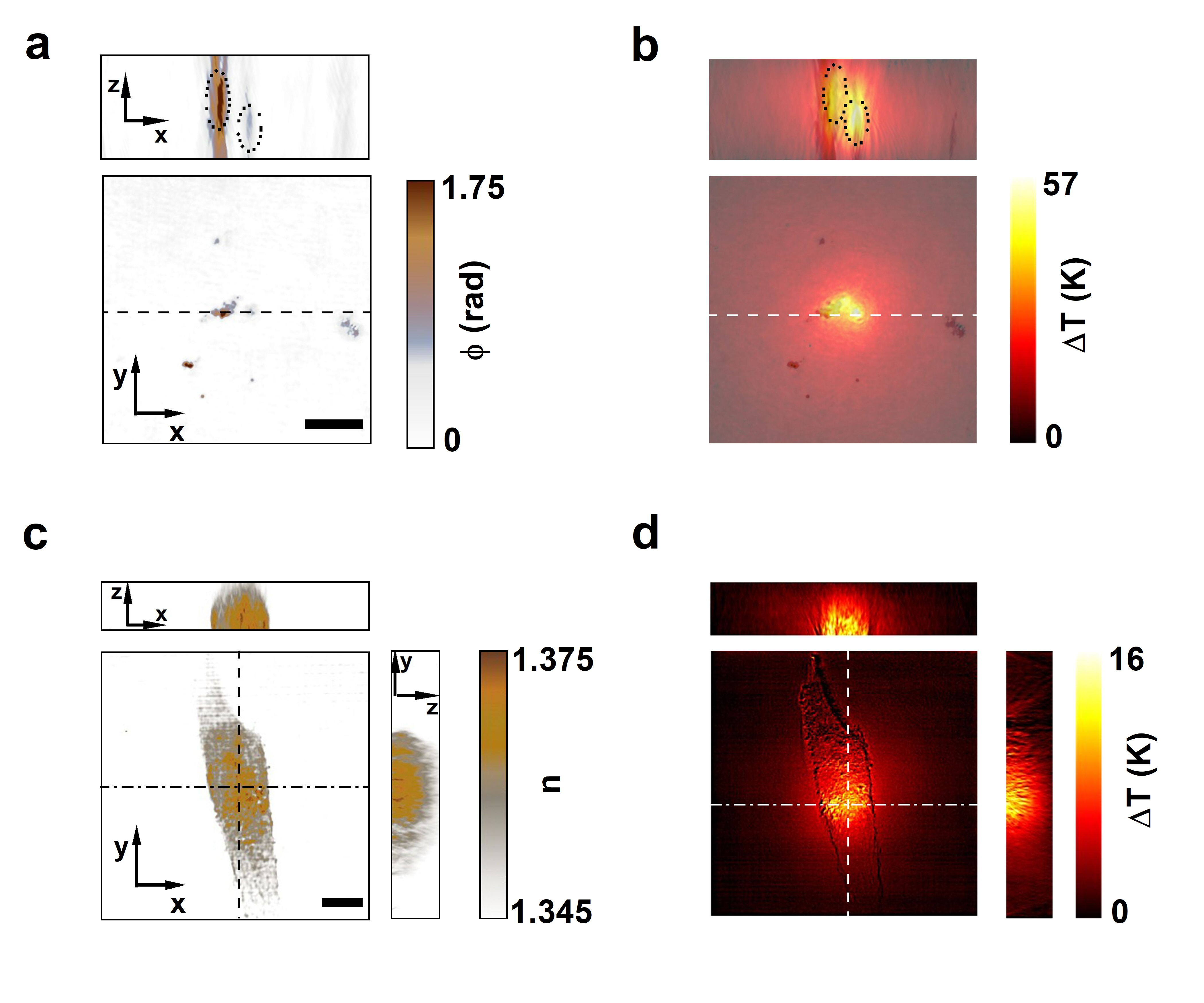}
    \caption{ \textit{Nanothermometry with spatially fixed 3D heat sources}. (a) 3D Phase image of nanoparticle clusters. The dashed circles act as a guide to the eye showing that the clusters are located in different planes. (b) Corresponding 3D temperature profile superimposed on the phase image when the nanoparticle cluster is excited. (c) 3D refractive index map of a fixed A549 lung cancer cell that has ingested AuNRs. (d) Corresponding 3D temperature profile when the cell was pumped with a 785 nm laser. The scale bars are 5 $\mu$m.  
} 
    \label{fig:Fig5}
\end{figure*}

So far we have studied thermal maps of AuNRs anchored to the surface of a glass substrate acting as a planar source of heat. A unique advantage of ODT is that it can measure temperature profiles originating from 3D heat sources, for instance from an ensemble of nanoparticle clusters distributed in a 3D matrix\cite{28}. To validate the versatility of ODT to non-planar heat sources, we immobilised nanoparticle clusters in polyacrylamide gels cast in microchambers of height 100 $\mu$m (SI section S1). Figure \ref{fig:Fig5}(a) shows the reconstructed phase image of the nanoparticle clusters within the sample volume obtained through digital hologram propagation \cite{16}. These nanoclusters were then excited using the same pump laser and Figure\ref{fig:Fig5}(b) shows the resulting thermal maps retrieved with ODT. The overlay of the 3D phase maps with the corresponding thermal map in figure\ref{fig:Fig5}(b) highlights the inherent property of ODT to colocalise the local temperature landscape with the 3D spatial distribution of complex objects.

To further understand the applicability of ODT thermometry to systems with refractive index inhomogeneities, we probed a A549 lung cancer cell that had previously internalised AuNRs. We quantified the increase in local temperature caused by photothermal conversion from the AuNRs inside the cell upon resonant excitation (see section S1 SI). Figure \ref{fig:Fig5} (c) and (d) show the 3D tomogram of a representative single cell, alongside the induced 3D thermal profile  upon irradiation with a 50 ms pump pulse respectively. Here the cell-internalized AuNRs act as non-planar heat sources embedded inside a complex refractive index environment, represented here by the cell. In this particular experiment, though the AuNRs were distributed throughout the volume of the cell, the much smaller illumination laser spot defined the size of the heat source. The spatial distribution inside the cell (particularly \textit{xz} cross-cut) shows that the temperature reached a maximum at a location away from the substrate; confirming that the system corresponds to a non-planar heat source (see section S6 SI).

\subsection{ODT thermometry: Nanorod colloids}
\begin{figure*}[h!]
    \centering
    \includegraphics[width=\linewidth]{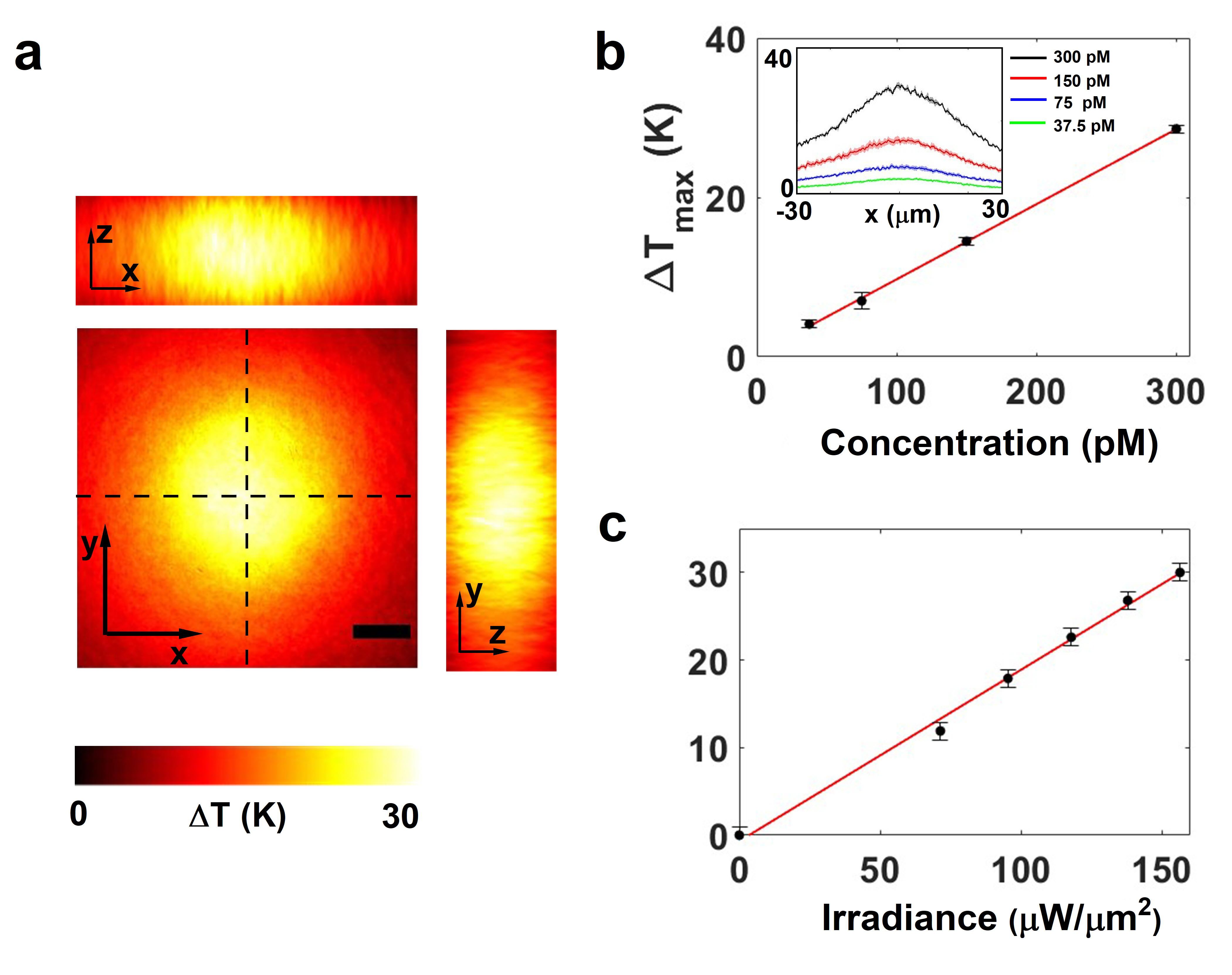}
    \caption{ \textit{Nanothermometry of Au nanorod colloids}. (a) 3D temperature of a colloidal sample in the microchamber of height 100 $\mu$m excited with a laser retrieved using ODT thermometry. The pump duration was kept at 20 ms and the frame rate at 10 Hz. \textit{Inset} represents the schematic of the temperature retrieval in colloidal nanoparticles using ODT. (b) The maximum rise in temperature as a function of the concentration of Au NRs in the chamber. The input irradiance was fixed at 151 $\mu W/\mu m^2$. \textit{Inset} shows the line profiles of the temperature plotted along y=0; z=0. (c) The maximum rise in temperature as a function of the input irradiance. The concentration of Au NRs was fixed at 300 pM. The scale bar is 5 $\mu$m.   
} 
    \label{fig:Fig6}
\end{figure*}

Apart from spatially fixed heat sources (AuNRs anchored on a glass substrate or clusters dispersed in poly-acrylamide gels/biological cells), we applied ODT thermometry to measure the temperature profile proceeding from dynamic environments. As a model system of this, we selected colloidal AuNR solutions. In detail, we prepared microchambers of height 100 $\mu$m filled with solutions of AuNRs of varying concentration, and probed them using ODT thermometry. In systems where the sources of heat can freely diffuse, large-scale spontaneous migration due to the formed temperature gradient, termed thermophoresis \cite{37}, represents a major bottleneck in the retrieval of temperature . When the nanoparticles are heated, they  move in response to the temperature gradient. However, the thermophoretic mobility depends on size and composition of the nanoparticles and also the duration of heating \cite{38,39,43,44}. To avoid large-scale thermophoresis in the sample we kept the pump pulse duration as 20 ms. We also tracked the phase shift induced in the probe beam across multiple pump cycles to ensure that there was a minimal change in the phase difference profile; thereby showing the heat source density per pump pulse did not vary drastically (figure S12, SI). 

Figure \ref{fig:Fig6} (a) shows the 3D thermal profile measured at AuNR concentration of 300 pM . To further characterize the temperature increase in colloidal nanoparticles, we systematically studied the temperature increase as a function of AuNR concentration and the input irradiance. Figures \ref{fig:Fig6}(b) and (c) depict the maximum temperature reached in the collodial system as a function of concentration and input irradiance respectively. The linear dependence of the maximum temperature both on the concentration (at constant irradiance, 151 $\mu$W/$\mu$m$^2$) as well as irradiance (at constant concentration, 300 pM) establishes the reliability of the temperature retrieval and shows that there is no large scale thermophoresis due to localized heating and thermal gradient.

Together these three proof-of-principle experiments establish the power and advantage of ODT thermometry over other existing methodologies. Namely, ODT delivers 3D thermal profile distributions from complex non-planar heat sources under transient thermodynamics regimes, i.e. not in steady-state. However, such advantages come at the cost of relatively more complex experimental setup and data processing steps as well as longer acquisition times compared to other QPI thermometries.

\subsection{Conclusions}
To summarize, we have studied experimentally and numerically the thermodynamics of a model non steady-state system consisting of an optically heated microchamber. We unraveled an important relationship between the time to reach steady-state and the chamber height and the thermal conductivity of the environment. We showed that ODT thermometry accurately retrieves the 3D temperature profile under nonsteady-state conditions and these results against numerical simulations and QPI thermometry. We showed the versatality of the ODT thermometry technique by applying it to systems with non-planar heat sources. We further presented a promising application of ODT thermometry by imaging the induced temperature profiles within biological cells upon plasmonic photothermal treatment. We also demonstrated its compatibility to retrieving temperatures from colloidal systems. We believe that the work presented here will have impact on multiple areas of research where accurate measurement of temperature is a key, such as photothermal therapy, photocatalysis, thermal lensing, and microfluidic optical traps. We also anticipate that the conclusions drawn in the article will stimulate further experimental and theoretical investigations on the development of more accurate and faster thermometry techniques, which will represent a significant step forward to better understanding heat-related processes at the nano and micro-scales.    

\section*{Acknowledgements}
The authors thank Helena Villuendas for their help in sample preparation and Guillaume Baffou for fruitful discussions.



\clearpage

\maketitle

\date{}
\maketitle
\fancypagestyle{firststyle}
{
   \fancyhf{}
\fancyhead[c]{Supporting Information}
  \fancyfoot[C]{S\thepage}
}

\thispagestyle{firststyle}

\pagestyle{fancy}
\fancyhf{}
\cfoot{S\thepage}


\appendix

\renewcommand{\thesection}{S\arabic{section}}
\renewcommand{\thefigure}{S\arabic{figure}}
\setcounter{figure}{0}
\setcounter{page}{1}
\section{Methods}
\subsection{Sample preparation}
Gold nanorods were synthesized using the method described in Nikoobakht et.al. \cite{48}. The glass substrate was uniformly coated with the synthesized gold nanorods using standard functionalization protocol described in detail in ~\citep{2}. The microchambers were, then, prepared by placing silicon gaskets of predetermined thickness (50 $\mu$m, 100 $\mu$m, 300 $\mu$m, and 500 $\mu$m) on the nanorod coated glass substrate. The gap in the silicon gasket was filled with $\sim$10 $\mu$l of DI water and the chamber was closed with a sapphire superstrate. The inner diameter of the silicon gasket was about 8 mm in all the cases. 

To immobilize nanoparticle clusters in polyacrylamide gels, we followed the standard operating protocol of preparing gels outlined by BioRAD \cite{54}. 

For nanothermometry experiments with biological cells, we used Lung cancer cells (ATCC CCL-185$^{TM}$) which had internalized gold nanorods. The protocol for the sample preparation is as follows. Petri dishes were cleaned in 70\% ethanol and sterilized in UV light for 10-20 min. We placed two silicon wells (gaskets) in a petri dish with a seeding concentration of about 5000 and 10000 cells/well. Then 500 $\mu$l of complete medium (Dulbecco's modified Eagle medium (DMEM) +10\% fetal bovine serum (FBS) +1\%penicillin–streptomycin (PS)) was added to the cell droplet and let the cells attach overnight. Then, the wells were washed with the medium without FBS. We added 4ml of 2nM gold nanorod solution to the wells and they were allowed to incubate overnight. The wells were then washed again with medium without FBS. Later, the cells were washed twice with PBS and incubated for 5 min in 4\% PFA and for 15 min in 2\% PFA respectively and kept in 1\% PFA. The microchamber was prepared by removing 1\% PFA and adding $\sim$10 $\mu$l of DI water and closing the chamber with another glass substrate on the top and gently applying pressure to fix the silicon wells.

\subsection{Experimental setup and Hologram processing}
\begin{figure}[H]
    \centering
    \includegraphics[width=\linewidth]{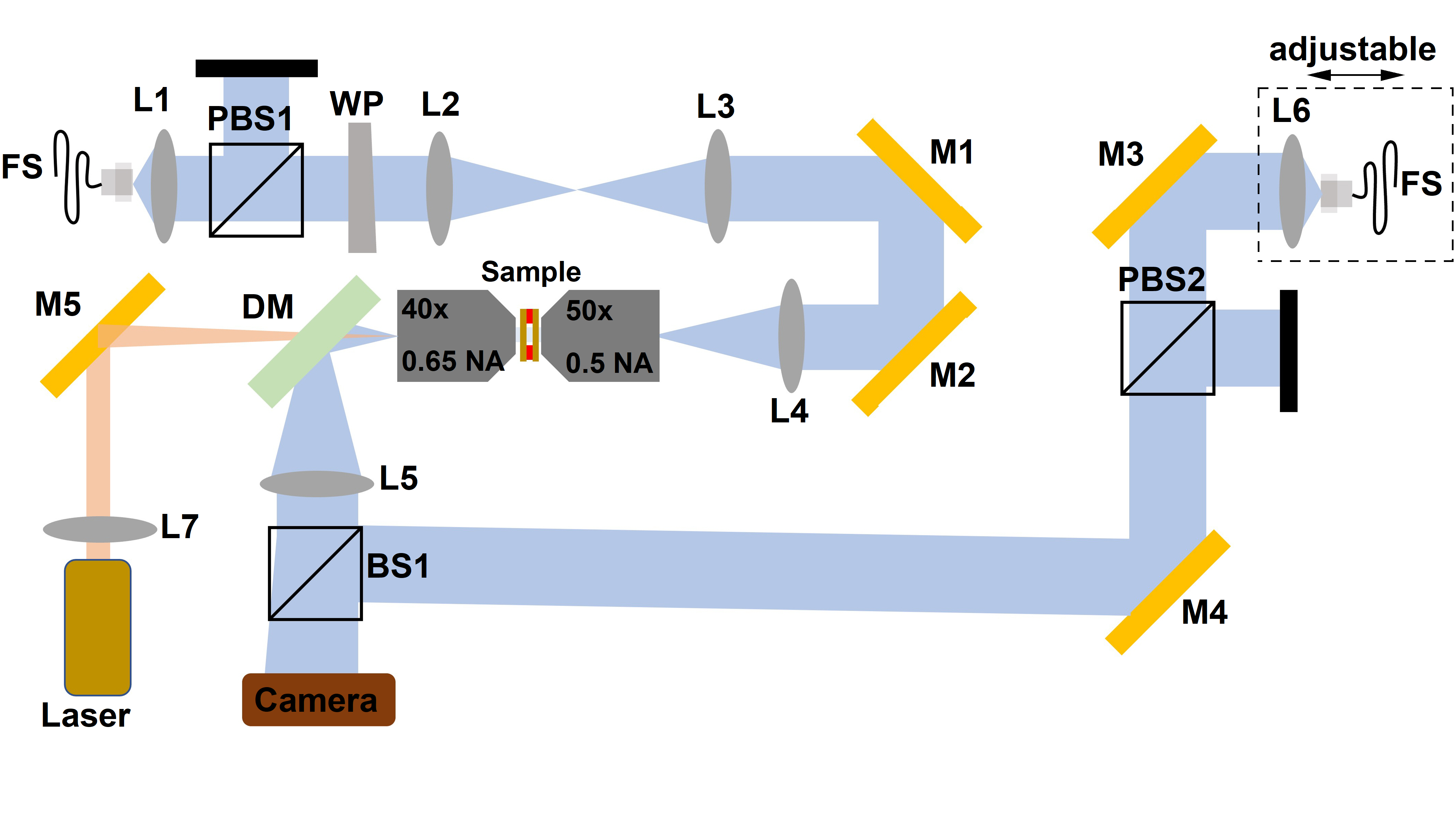}
    \caption{Schematic of the experimental setup. \\ \textbf{Legend:} \textit{ L-Lens, M-mirror, PBS-Polarizing beam splitter, FS-Fiber source, BS-Beam splitter, DM-Dichroic mirror}.}
    \label{fig:FigS1}
\end{figure}
Figure~\ref{fig:FigS1} shows a schematic of the experimental setup used to measure phase maps in off-axis holography configuration in pump-probe manner. The probe laser (465 nm) was split into reference and object beams using a fiber splitter. In the object path, the probe laser was focused onto the back aperture of the objective lens (50x, 0.5NA) to generate a wide field illumination using a combination of lenses L2 ($f_{L2}$=30 mm), L3($f_{L3}$=150 mm), and L4 ($f_{L4}$=150 mm). The probe light was collected in transmission configuration by a 40x 0.65 NA objective lens and the collected light was projected onto the camera using lens L5 ($f_{L5}$=250 mm) creating a magnification of 55 ($\frac{f_{L5}}{f_{objective}}$). The angle of illumination at the sample plane was controlled by the wedge prism WP. The reference beam was projected onto the camera at a small angle with respect to the optic axis of the microscope with the help of mirrors M3 and M4. The polarization of both object and reference path was fixed using polarizing beam splitters PBS1 and PBS2. The path length of the reference beam  was adjusted by placing the fiber source module on an adjustable stage so as to match that of the object path. The gold nanorods were excited using a pump beam of wavelength 785 nm through 40x objective lens. The pump laser was not expanded and a long focal length lens, L5 ($f_{L4}$=500 mm), was used to focus the pump laser onto the back aperture of the objective lens to generate a spot size of $\sim$10 $\mu$m. An FPGA card was used to synchronize the pump and probe lasers to the camera acquisition. The frame rate of the camera was set at 10Hz for all experiments and the pump and probe pulse duration was varied according to the requirements. 

The measured hologram were processed by first taking the Fourier transform which revealed real, twin, and zero-order images in the $k$-space. The real image was filtered using a hard aperture selection followed by frequency demodulation\cite{49}. The demodulated image was then inverse Fourier transformed to get the complex electric field, of which the real part provided the amplitude image and the imaginary part the phase image. This step was repeated for all angles of illumination.  

For all angles of incidence, we measured the phase maps for pump ON state and OFF state to generate a phase difference image. This step was repeated multiple times (50 phase images) and the resulting phase difference maps were averaged to increase the signal to noise ratio and improve the phase sensitivity.

\section{Thermal Imaging using optical diffraction tomography (ODT)}
\begin{figure}[H]
    \centering
    \includegraphics[width=\linewidth]{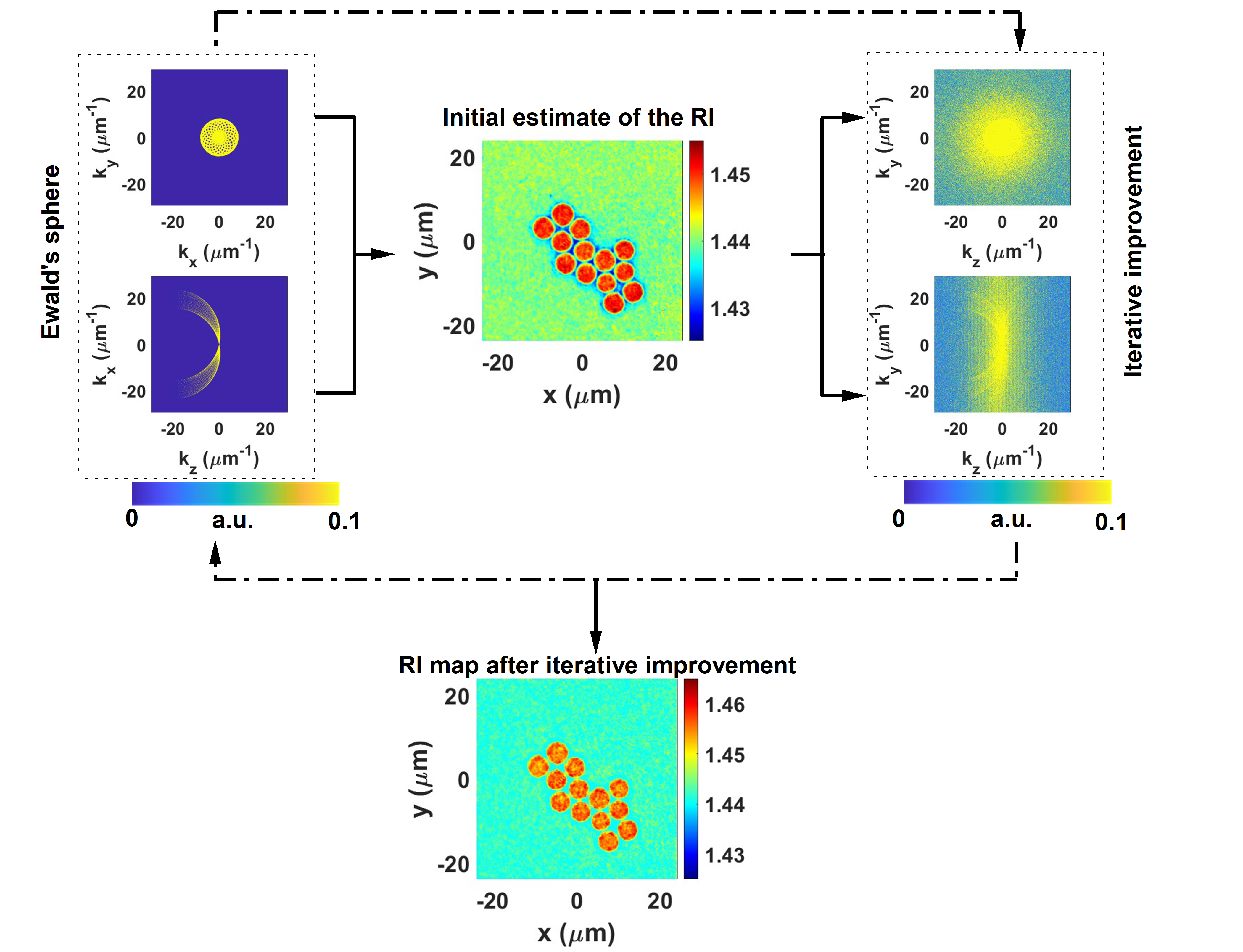}
    \caption{The process flow in ODT explaining the reconstruction of 3D refractive index profile of silica beads dispersed in PDMS matrix. }
    \label{fig:FigS2}
\end{figure}
Optical diffraction tomography (ODT) has been widely used as an imaging tool to map the refractive index in 3D. ODT relies on calculating refractive index profile from multiple phase and amplitude images, measured by changing the angle of illumination, using the Fourier diffraction theorem. According to the said theorem, for any refractive index composition n(r) immersed in a medium of refractive index $n_m$ we can write
\begin{equation}
    \hat{F}(K_x,K_y,K_z)=\frac{ik_z}{\pi}\hat{U_s}(k_x,k_y;z=0)
    \label{eq1}
\end{equation}
where $\hat{F}$ is the 3D Fourier transform of the object function $f=-\frac{2\pi n_m}{\lambda^2}((\frac{n(r)}{n_m})^2-1)$ with $\lambda$ as wavelength of illumination. $\hat{U}$ is the 2D Fourier transform of the scattering wave calculated using Rytov approximation. According to the Rytov approximation, we can express the scattering wave ($U_s$) as $U_s(x,y)$=$ln(\frac{U(x,y)}{U_{back}(x,y)})$ with $U(x,y)$ and $U_{back}(x,y)$ as the retrieved complex electric field in the presence of the sample and the background electric field respectively. In the case of ODT thermometry, $U(x,y)$ and $U_{back}(x,y)$ correspond to the complex electric fields measured with pump ON and pump OFF respectively. $k_z$ is related to the lateral spatial frequencies $(k_x,k_y)$ as $\sqrt{(\frac{2n_m\pi}{\lambda})^2-k_x^2-k_y^2}$. For each illumination angle the spatial frequencies of the incident wave vector changes and so does $(K_x,K_y,K_z)$. Thus, one can map different regions of the $k$-space of the object function with measuring multiple 2D complex electric field images by changing the angle of incidence. Thus mapped 3D object in the $k$-space is usually called as an \textit{Ewald's sphere}.  Finally, the inverse Fourier transform of $\hat{F}$ yields the 3D profile of the object function in real space which can readily be translated to the 3D refractive index profile. 

It is important to note that due to the limited numerical aperture of the collection objective lens it is not possible to fill the whole \textit{Ewald's sphere} through experimental data. This problem reflects is usually referred to as the \textit{missing cone} problem. There are multiple methods to solve the \textit{missing cone} problem among which constrained iterative regularization is widely used. For ODT thermometry the missing points on the \textit{Ewald's sphere} were filled satisfying two physical constraints
\begin{itemize}
    \item The refractive index of water at an elevated temperature can not be greater than that at an ambient temperature, as water has a negative thermo-optical co-effecient.
    \item The refractive index of the glass substrate is constant as the heat induced refractive index change in glass is negligible under experimental conditions.
\end{itemize}

Then the 3D refractive index maps were transformed to thermal maps using an empirical equation:
\begin{equation}
    n(T)=\sum_{j=0}^{P} b_jT^j
    \label{eq2}
\end{equation}
where $T$ is temperature and $b_j$s are expansion co-efficients. We consider upto P=4 with values given as, \cite{50}\\
$b_0$=1.34359\\
$b_1$=-1.0514x10$^{-4}$\\
$b_2$=-1.5692x10$^{-6}$\\
$b_3$=5.7538x10$^{-9}$\\
$b_4$=-1.2873x10$^{-11}$\\

\subsection{Benchmarking ODT}
As a control experiment we benchmarked the ODT technique by measuring refractive index of silica microparticles dispersed in PDMS matrix. In this case $U(x,y)$ was the complex electric field maps measured in the presence of the particle and $U_{back}(x,y)$ was the background electric field maps of bare PDMS matrix (without particles) (see equation\ref{eq1}). The initial estimation of refractive index of microparticles was improved using the iterative regularization step with constraint as: the refractive index of beads was greater than the refractive index of PDMS. Figure \ref{fig:FigS2} shows the schematic of the tomographic reconstruction of the refractive index profile of silica microparticles dispersed in PDMS matrix and also indicates the effect of iterative regularization. 

\subsection{Phase and temperature sensitivity}
\begin{figure}[H]
    \centering
    \includegraphics[width=\linewidth]{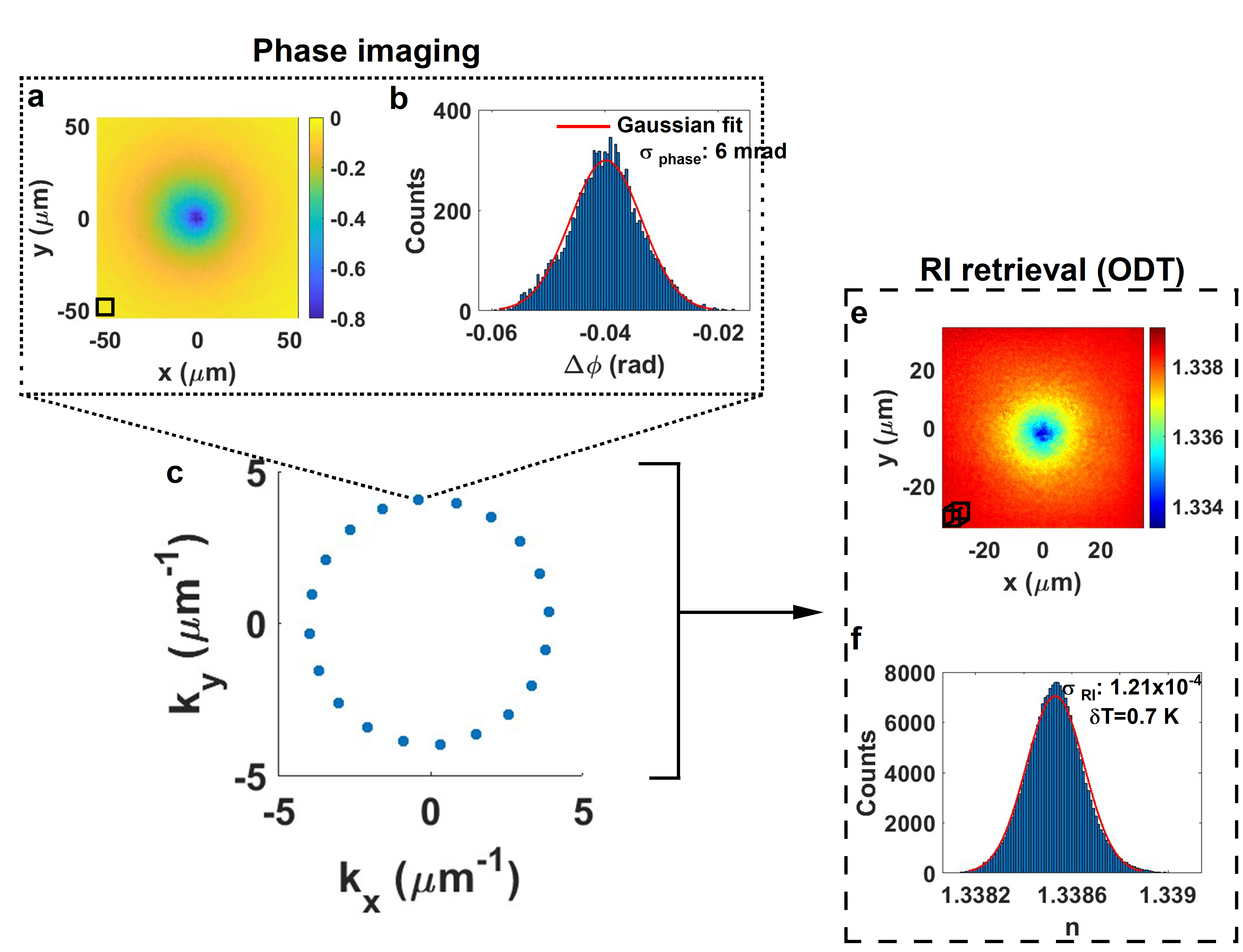}
    \caption{(a) A representative phase difference map measured for a single angle of incidence. (b)Distribution of background phase values inside the black square shown in (a). An refractive index map was constructed by using twenty phase difference maps for different angles of incidence as shown in (c). (d) Corresponsing refractive index map and (e) Distribution of background refractive index values inside the black cube shown in (e). }
    \label{fig:FigS3}
\end{figure}

To further characterize the sensitivity of thermal imaging using ODT technique, we performed statistical analysis of the phase map as well as the retrieved refractive index map. The phase senstivity, which is the minimum distinguishable phase change, was calculated by plotting the histogram of values in the background area (inside a square defined along the corner of the image) of the phase map as shown in figures \ref{fig:FigS3}(a) and (b). The histogram values were fit to a gaussian distribution and the sensitivity, $\sigma_{phase}$, was found to be 6 mrad. In a typical RI tomogram we use 20 of such phase maps by changing the input angles as shown in Figure \ref{fig:FigS3}(c). The RI sensitivity was calculated in a similar manner as that of the phase and was found to be $\sigma_{RI}$=1.21x10$^{-4}$ which corresponds to a temperature sensitivity of $\delta T$=0.7K (see figures \ref{fig:FigS3} (e) and (f)).  

\section{Thermal imaging using QPI }
The algorithm to retrieve temperature map from an optical phase difference map has been outlined in detail in ref\citep{4}. The measured phase difference map was converted into optical path difference (OPD) map. Then, the temperature retrieval was done in three steps:
\begin{enumerate}
    \item  First we assume that the system has reached the steady state distribution of the form $\frac{P_0}{r}$, where $P_0$ is the absorbed power and $r$ is the radial coordinate. Then we define the Green's function for OPD as $G_{OPD}=\beta_1sinh^{-1}(\frac{h}{\sqrt{x^2+y^2}})$ where $h$ is the height of the microchamber and $\beta_1$ corresponds to the first order thermo-optical co-efficient of water. We estimated the heat source density (HSD) by denconvolution of the OPD map and the $G_{OPD}$ using Tikhonov deconvolution method. 
    \item  We then convolved the obtained HSD map with the Green's function for Laplace equation, $G_{Th}=\frac{1}{4\pi \kappa_{water}r}$ where $\kappa_{water}$ is the thermal conductivity of water, to get an initial estimate of temperature.
    \item  To consider the higher order thermo-optical co-effecients, we iteratively minimized the difference between the measured OPD map, $l_0$(x,y), and a path difference map calculated using $\int_{0}^{h} \delta n(x,y,z) dz$, where $\delta n(x,y,z)$ is the change in refractive index of water calculated using the retrieved temperature map by considering higher orders (up to 4) of thermo-optical co-effecients of water (see equation\ref{eq2}) .
\end{enumerate}
For thermal imaging using QPI, we averaged twenty different phase maps measured by changing the angle of incidence in a manner similar to the building of \textit{Ewald's sphere}, but in 2D. The resulting OPD map was employed in the temperature retrieval.

\subsection{Effect of chamber height on temperature retrieval}
\begin{figure}[H]
    \centering
    \includegraphics[width=\linewidth]{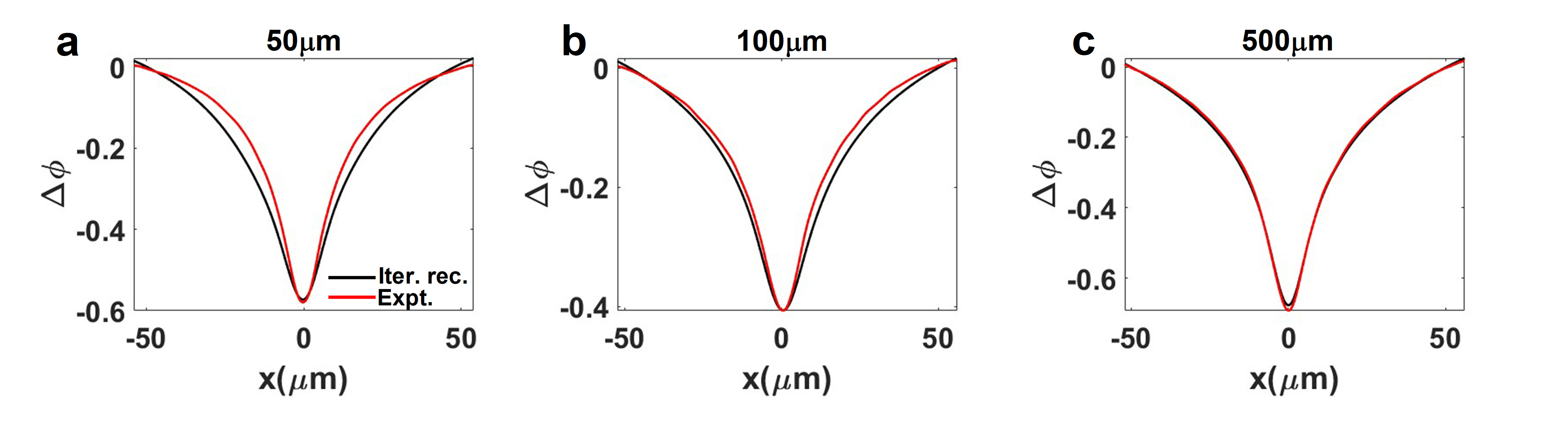}
    \caption{The effect of chamber height, (a) 50 $\mu$m (b) 100 $\mu$m (c) 500 $\mu$m, on the convergence of the iterative deconvolution procedure to calculate the temperature using QPI approach.  }
    \label{fig:FigS4}
\end{figure}

  As explained in the previous subsection, the retrieval of the temperature map is based on the assumption that the temperature has reached a steady state and is of the form $\frac{P_0}{r}$. This is strictly true only in the case where the medium of heat transfer is quasi-infinite in nature. However when the height of the chamber is relatively small one has to take the thermal properties of the superstrate into account as discussed in ref\citep{5}. However, it is tricky to obtain a closed form expression for HSD through deconvolution of the OPD maps if one considers the complete 3 layer model. The deviation from the quasi infinite model in temperature gets reflected in the iterative minimization of OPD error (step 3 in the temperature retrieval) as shown in figure \ref{fig:FigS4}. The deviations of the reconstructed OPD map from the measured one decereases as one increases the chamber height from 50 $\mu$m to 500$\mu$m.

\section{Temperature transients}
\begin{figure}[H]
    \centering
    \includegraphics[width=\linewidth]{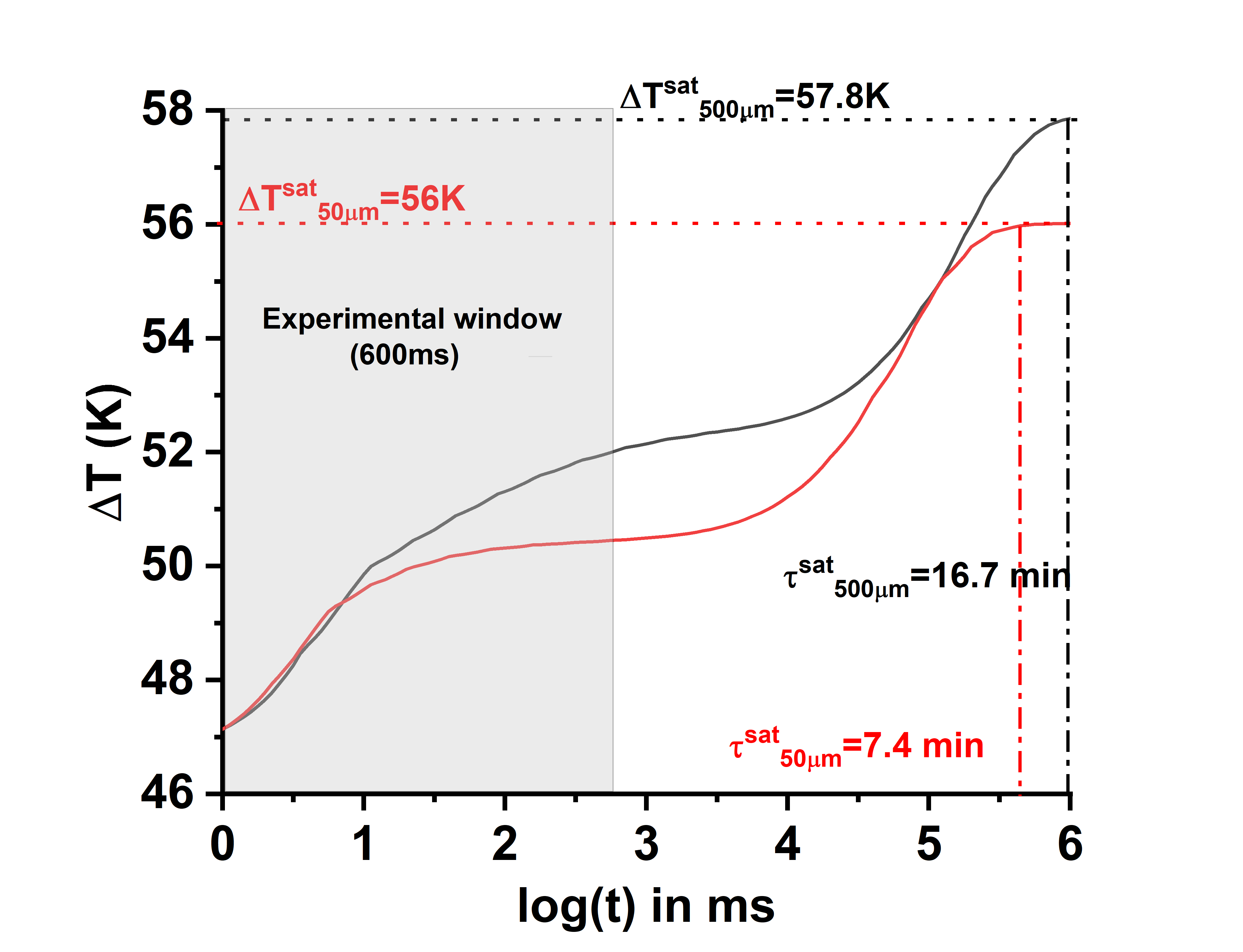}
    \caption{Numerically calculated time evolution of temperature for a microchamber of height 500 $\mu$m and 50 $\mu$m keeping the heating beam size constant.}
    \label{fig:FigS5}
\end{figure}

Time taken to reach steady state for a quasi-infinite system is defined by its diffusivity, $a_s$, and the diameter of the heat source, $D$, and is of the form $\frac{D^2}{4a_s}$.  Most of the experimental systems, eg.\ microchambers, are far from being quasi-infinite systems because of their limited chamber height. In such cases, the temperature dynamics depend crucially on the chamber height. To understand the temperature dynamics in microchambers we performed finite element method (FEM) based numerical simulations using COMSOL multiphysics. The lateral diameter of the chamber was fixed to 10 mm and the height of the chamber was varied. The substrate was defined as glass ($\kappa_{glass}$=0.9 W/mK) and the superstrate as sapphire ($\kappa_{saph}$=30W/mK). To mimic the experimental situation the chamber was heated by a gaussian heat source with a diameter of 10 $\mu$m. Natural air convection was assumed at the boundaries (both glass side and sapphire side) as in the case of experiments. Figure \ref{fig:FigS5} shows numerically calculated evolution of maximum temperature as a function of pump duration for chamber heights of 50 $\mu$m and 500 $\mu$m. For both cases the time taken to reach steady state was significantly different from the values predicted by $\frac{D^2}{4a_s}$ i.e., 7.4 min for the chamber of height 50 $\mu$m and 16.7 min for the chamber of height 500 $\mu$m. 

A striking feature to note is that the temperature reaches an intermediary plateau, a quasi steady state, before reaching the actual steady state. This can be understood by considering two thermalizing regimes of the system: \textit{local} and \textit{global}. When the ensemble of nanoparticles were heated in a thermodynamically closed system like a microchamber, the heated nanoparticles thermalize fast with its local environment and reaches an intermediary steady state which is termed as the \textit{local} thermalization regime (the plateau in figure \ref{fig:FigS5}). The rate at which the entire microchamber thermalized was determined, primarily, by the natural convection at the boundaries. This \textit{global} thermalization happened at a slower pace and at a higher temperature which was a function of the volume of liquid inside the chamber. Such temperature evloution has been observed in the past in the case of multi nanoparticle heating \cite{53}. It has to be noted that the conclusions drawn here strictly apply to the case where natual air convection is the mechanism with which the microchambers interact with the surrounding environment. If the substrate is in contact with any kind of thermal sinks eg., metallic case of the objective lens, the dynamics of the system will be altered. 

\begin{figure}[H]
    \centering
    \includegraphics[width=0.5\linewidth]{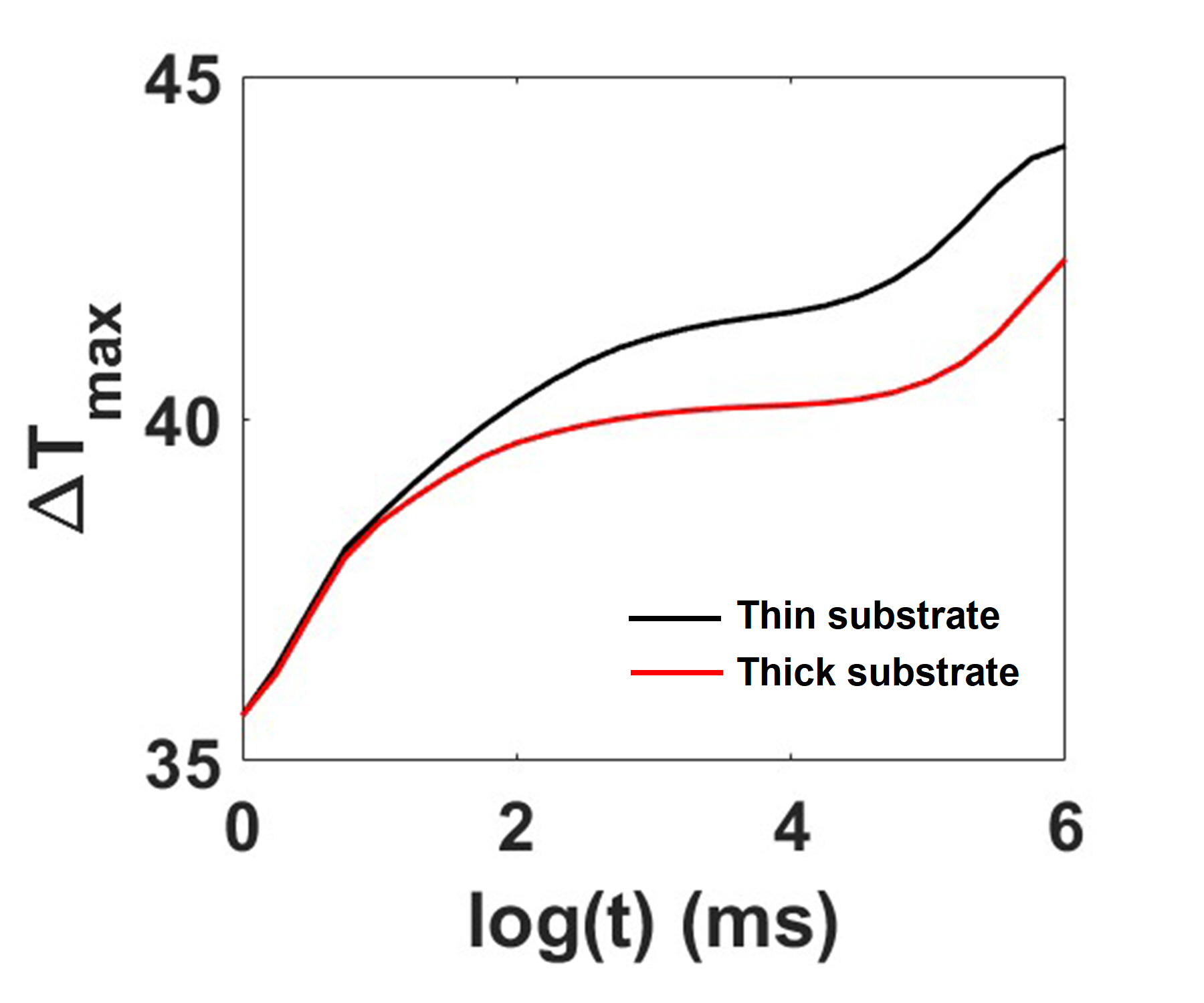}
    \caption{Numerically calculated time evolution of temperature profile for a microchamber of height 500 $\mu$m with thick substrate (1 mm) and thin substrate (100 $\mu$m).  
    }
    \label{fig:FigS12}
\end{figure}

To understand the effect of substrate thickness on the temporal evolution of the thermal profile, we performed numerical calculations considering two different substrate thicknesses: 100 $\mu$m and 1 mm. Figure \ref{fig:FigS12} shows the calculated thermal profiles of a microchamber of height 500 $\mu$m with 1 mm and 100 $\mu$m thick substrates respectively. The beam size was kept at 10 $\mu$m. The microchamber with thinner substrate reaches the steady state faster albeit at a higher temperature compared to a chamber with a thicker substrate. The increase in the time to reach steady state for the microchambers with thicker substrate was due to the increase in the volume of the material to thermalize compared to the chambers with the thin substrate. As the sources of heat were located in glass - water interface, the thicker substrate introduces a longer path in glass for the heat to diffuse and decay. This makes the effect of air convection (heat accumulation in the chamber) comparatively lesser than the thin substrate case.

\subsection{Effect of superstrate on temperature dynamics}
\begin{figure}[H]
    \centering
    \includegraphics[width=\linewidth]{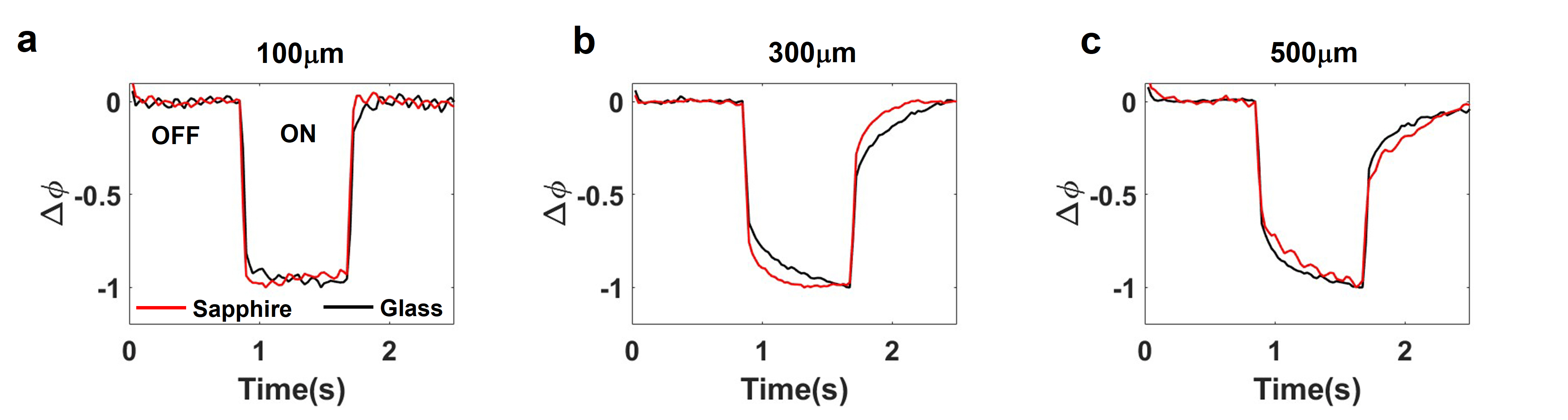}
    \caption{The effect of sapphire superstrate on the phase evolution of the microchambers for chamber heights (a) 100 $\mu$m , (b) 300 $\mu$m, and (c) 500 $\mu$m.}
    \label{fig:FigS6}
\end{figure}

To understand the effect of the thermal conductivity of the superstrate on temperature dynamics we probed microchambers with glass and sapphire as superstrates keeping glass as the substrate. Figure \ref{fig:FigS6} shows the evolution of normalized phase accumulation ($\Delta\phi_{max}$) as a function of three different chamber heights. For both 100 $\mu$m and 500 $\mu$m chambers the phase evolution was almost independent of the superstrate while there was a marginal change for a 300 $\mu$m chamber. For the chamber height of 100 $\mu$m, the volume of liquid itself is small enough and introduction of sapphire as the superstrate does not quicken the temporal evolution of temperature, atleast within the experimental setting. Also, if the superstrate is very far (500 $\mu$m case) to the heat source the thermodynamics is mostly defined by the natural convection and thermal conductivity of the superstrate does not have a major role to play. However with an intermediary height (300 $\mu$m), sapphire plays a role in quickening the time evolution by acting as an efficient heat sink.       
\subsection{Effect of thermal conductivity of the surroundings on the temperature dynamics}
  \begin{figure}[H]
    \centering
    \includegraphics[width=0.65\linewidth]{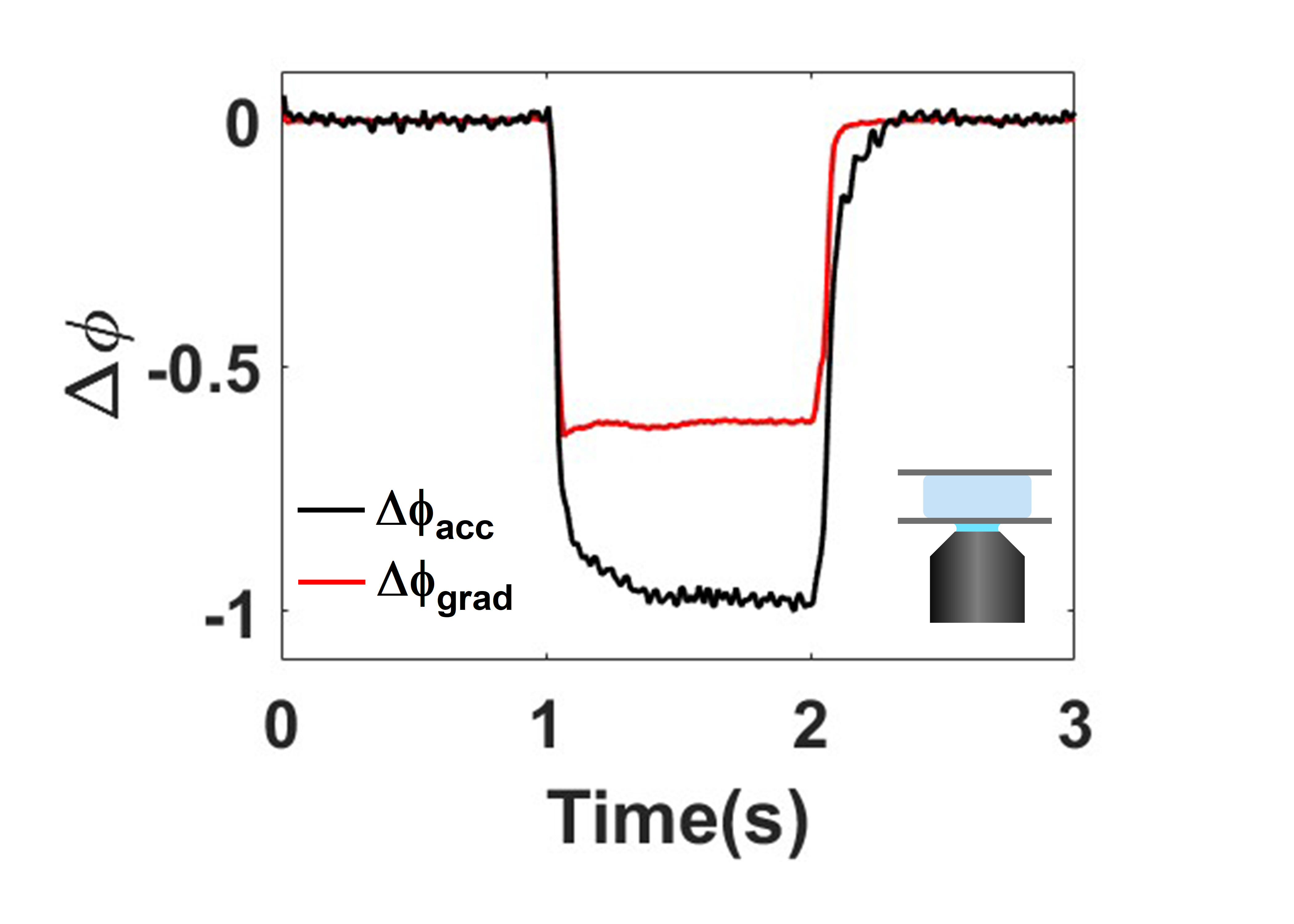}
    \caption{Evolution of phase difference for a 500 $\mu$m chamber probed using an oil immersion objective lens.}
    \label{fig:FigS11}
\end{figure}

Figure \ref{fig:FigS11} shows the temporal evolution of phase gradient ($\Delta\phi_{grad}$) and maximum phase accumulation ($\Delta\phi_{acc}$) for a microchamber of height 500 $\mu$m when probed using an oil immersion objective lens for a pump duration of 1 s. The phase accumulation saturates within the time duration of the experiment while the phase gradient shows an oscillatory behaviour before saturating (similar to figure 2). 

\section{Non-steady state thermometry of planar heat sources with ODT}
  \begin{figure}[H]
    \centering
    \includegraphics[width=\linewidth]{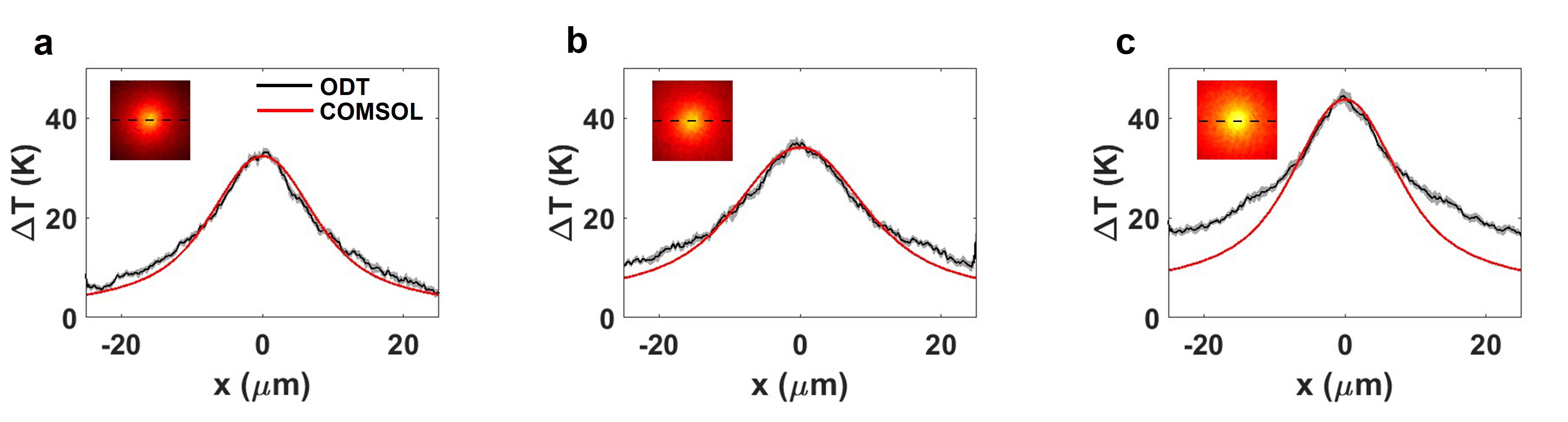}
    \caption{\textit{\textbf{ODT} thermometry, the effect of pump pulse duration}. Comparison of the line profile of the temperature along Y=0 for a chamber of height 500 $\mu$m pump duration (a) 5 ms, (b) 20 ms, and (c) 80 ms respectively with the numerically calculated thermal profile. Insets represent the XY crosscuts of the temperature profile in each case.}
    \label{fig:FigS8}
\end{figure}

To better understand the deviation of the thermal profile of the microchamber of height 500 $\mu$m from the numerically calculated one, we probed the microchamber as a function of pump pulse durations. Figures \ref{fig:FigS8} (a) - (c) show the comparison of line profiles of temperature as retrieved by ODT thermometry and numerical calculations using COMSOL for a pump pulse duration of 5 ms, 20 ms, and 80 ms respectively. The camera frame rate was fixed at 10 Hz. The thermal profiles retrieved by ODT match very well with the numerical simulations for a pump pulse duration of 5 ms and 20 ms as there was minimal residual heat interference in the pump-probe cycle. However, in the case of an 80 ms pump pulse, there is a considerable interference of the residual heat resulting in the deviation of the thermal profile from the numerically calculated one.

  \begin{figure}[H]
    \centering
    \includegraphics[width=\linewidth]{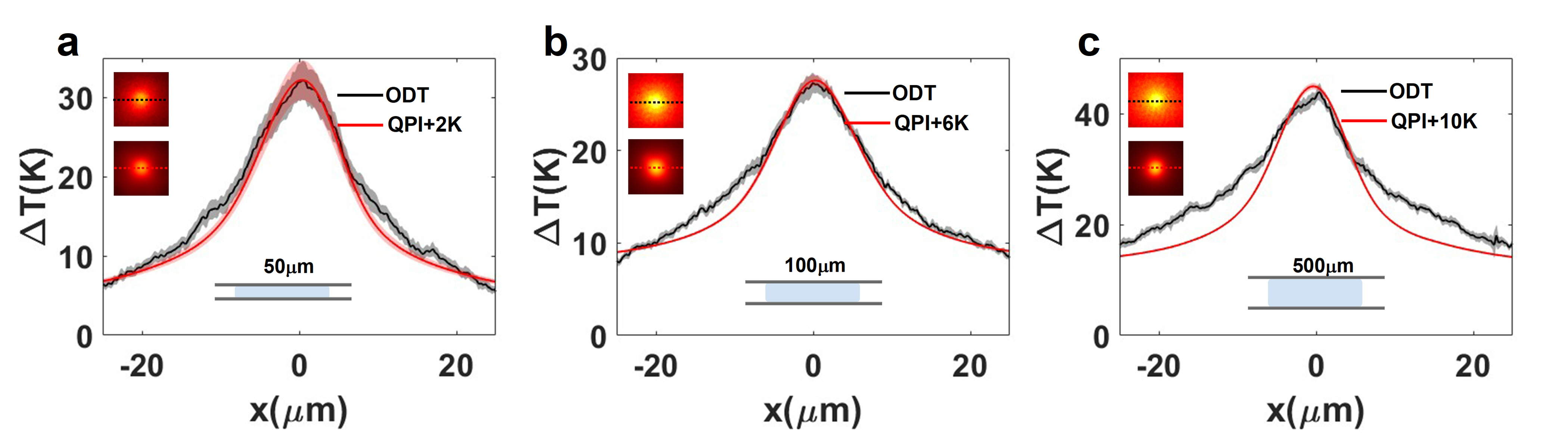}
    \caption{\textit{\textbf{ODT} vs \textbf{QPI}, the effect of chamber height}. Comparison of the line profile of the temperature along Y=0 for chamber heights (a) 50 $\mu$m, (b) 100 $\mu$m , and (c) 500 $\mu$m respectively. The pump duration was kept at 80 ms and the camera rate at 10 Hz. Insets represent the XY crosscuts of the temperature profile in each case.}
    \label{fig:FigS7}
\end{figure}

We utilized QPI thermometry to benchmark the thermal profiles retrieved by ODT technique.  In the case of microchambers, due to the accumulation of heat, the system experiences a constant increase in the thermal floor keeping the shape of the thermal profile the same, thus pushing the system out of the steady state. As mentioned in the previous sections, QPI thermometry is based on two important assumptions: (i) The system has reached the steady state and the temperature profile follows a $\frac{1}{r}$ decay. (ii) The sources of heat are all located in one plane. Figures \ref{fig:FigS7} (a)-(c) show the comparison of the line profile of the temperature along Y=0 for microchamber of height 50 $\mu$m, 100 $\mu$m, and 500 $\mu$m respectively. The pump pulse duration was kept at 80 ms and the camera frame rate at 10 Hz. A constant value was added to the temperature profiles retrieved by QPI for better visualization. The shape thermal profiles retrieved by ODT technique matched very well with QPI except for the constant value of the thermal floor, as predicted by the phase imaging, showing the microchamber was out of steady state.

\section{Thermal imaging of biological cells}
  \begin{figure}[H]
    \centering
    \includegraphics[width=\linewidth]{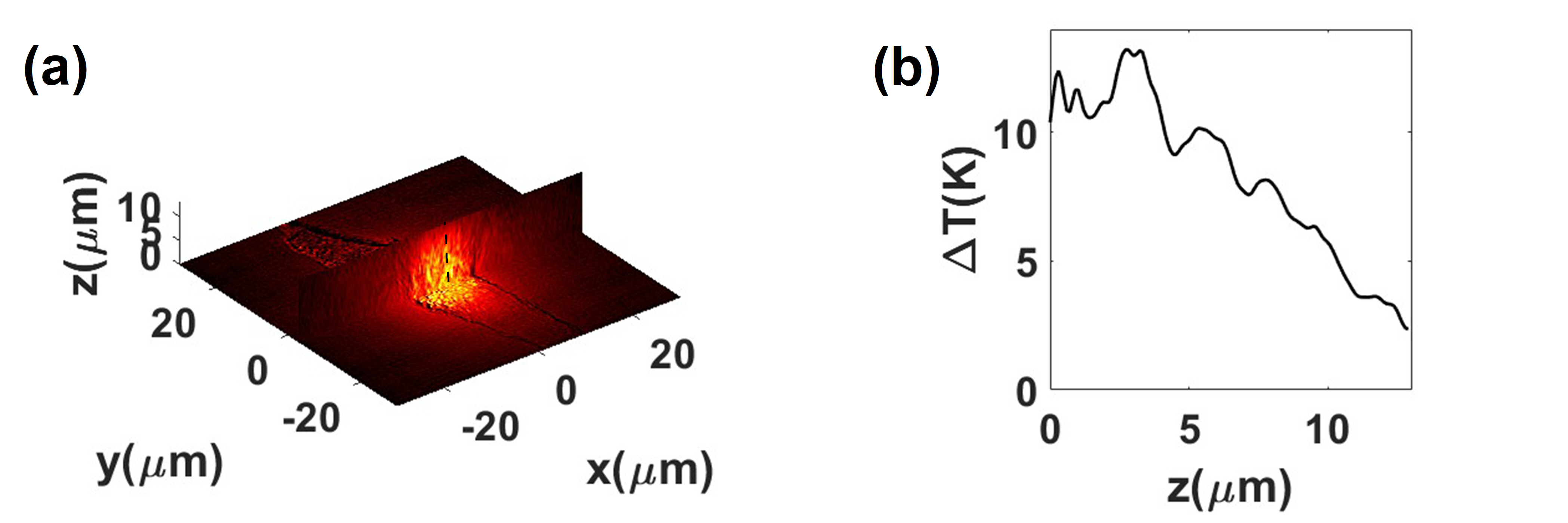}
    \caption{(a) 3D temperature plot of Au NR ingested A549 lung cancer cell excited by a 785 nm laser. (b) Corresponding line profile of the temperature along the dashed line shown in (a) }
    \label{fig:FigS9}
\end{figure}

To further confirm the non-planar nature of Au NRs ingested by the biological cell, we plot a line profile of the temperature along the axial direction as shown in figure\ref{fig:FigS9}. We can see that the distribution of temperature reached the maximum value away from the substrate (z=0). Additionally, the line profile deviated from the $\sim \frac{1}{r}$ decay from the substrate unlike a planar heat source.

\section{ODT thermomerty of Au nanorod colloids}
  \begin{figure}[H]
    \centering
    \includegraphics[width=\linewidth]{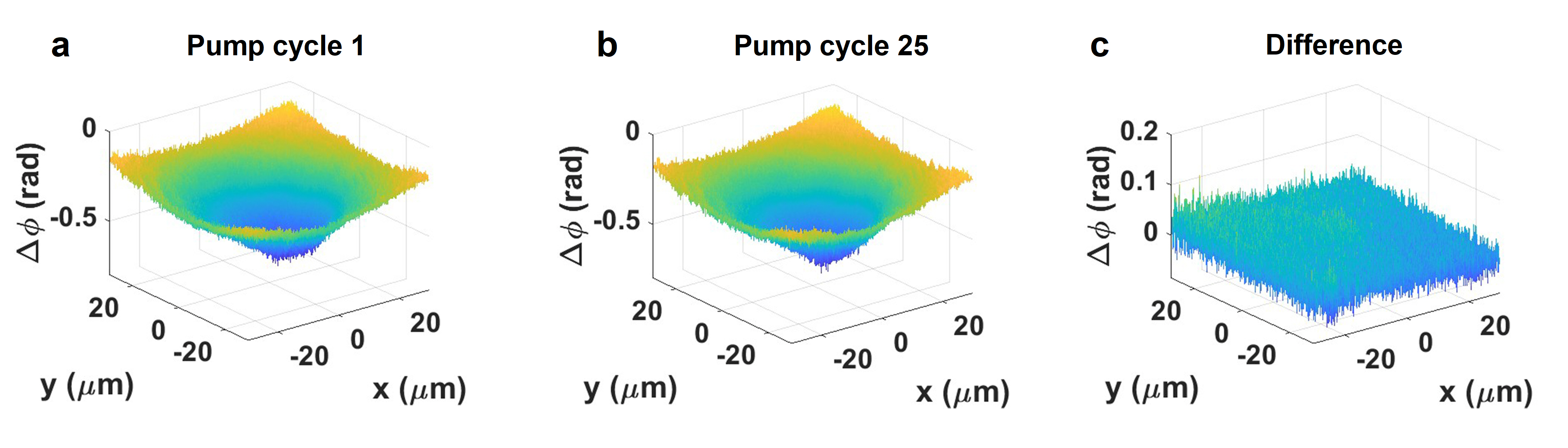}
    \caption{(a,b) Experimentally measured phase difference images for pump cycles 1 and 25 respectively. (c) The difference of the phase images for pump cycles 1 and 25.   }
    \label{fig:FigS10}
\end{figure}

The main concern in the thermometry of colloids is the large-scale movement of the nanoparticles due to the generated temperature gradient, termed thermophoresis. To avoid thermophoresis, we kept the pump duration to 20 ms. We further investigated the phase difference map for different pump cycles. The phase difference map gives an estimate of the heat source density generated per pump cycle. Figure \ref{fig:FigS10} shows the measured phase difference map for pump cycles 1 and 25 as well as the calculated difference between them. Notwithstanding multiple pump-probe cycles, the phase difference maps did not alter considerably demonstrating that the heat source density generated per pump pulse remained the same.

\bibliography{ref}
\bibliographystyle{naturemag}

\end{document}